\documentclass[a4paper,12pt]{article}
\usepackage{amssymb}
\usepackage{wick}
\usepackage{natbib}
\usepackage{mathbbol}
\usepackage{colortbl}

\title{%
{\vspace{-1cm}\bf Notes on non-trivial and logarithmic CFTs with \boldmath{$c=0$}}
}
\author{%
  {\normalsize\sc Michael Flohr\thanks{{\tt 
    flohr@th.physik.uni-bonn.de}}}\ \ \ \ {\normalsize and}\ \ \
  {\normalsize\sc Annekathrin M\"uller-Lohmann\thanks{{\tt 
    anne@th.physik.uni-bonn.de}}}\\[0.5cm]
  {\normalsize\slshape Physikalisches Institut}\\[-0.1cm]
  {\normalsize\slshape University of Bonn}\\[-0.1cm]
  {\normalsize\slshape Nussallee 12}\\[-0.1cm]
  {\normalsize\slshape D-53115 Bonn, Germany}
}
\date{{\small\today}}

\def\bra{\langle}
\def\ket{\rangle}

\def\vev#1{\langle#1\rangle}
\def\vvv#1#2{\langle#1|#2\rangle}
\def\nop#1{\mbox{:$#1$:}}
\def\id{\mathbb{I}}
\def\lid{\tilde\mathbb{I}}

\newcommand{\be}{\begin{equation}}
\newcommand{\ee}{\end{equation}}
\newcommand{\bea}{\begin{eqnarray}}
\newcommand{\eea}{\end{eqnarray}}
\newcommand{\s}[0]{\sigma }
\newcommand{\e}[0]{\varepsilon }
\newcommand{\tpz}[0]{\theta^+ (z)}

\newcommand{\tpw}[0]{\theta^+ (w)}
\newcommand{\dtpw}[0]{\partial \theta^+ (w)}

\newcommand{\dddtpw}[0]{\partial^3 \theta^+ (w)}
\newcommand{\ddddtpw}[0]{\partial^4 \theta^+ (w)}

\newcommand{\tmw}[0]{\theta^- (w)}
\newcommand{\dtmw}[0]{\partial \theta^- (w)}

\newcommand{\dddtmw}[0]{\partial^3 \theta^- (w)}
\newcommand{\ddddtmw}[0]{\partial^4 \theta^- (w)}

\newcommand{\lzw}[0]{\log (z-w)}

\newcommand{\ctpztmw}[0]{\wick{1}{<1 \tpz >1 \tmw}}

\newcommand{\cons}[0]{\mathrm{i}\sqrt{2} \alpha_0}
\newcommand{\eins}[0]{\lid }
\newcommand{\del}[0]{\partial }
\newcommand{\pz}[0]{\phi (z)}
\newcommand{\dpz}[0]{\partial \phi(z)}
\newcommand{\ddpz}[0]{\partial^2 \phi(z)}
\newcommand{\pw}[0]{\phi (w)}
\newcommand{\dpw}[0]{\partial \phi (w)}
\newcommand{\ddpw}[0]{\partial^2 \phi (w)}
\newcommand{\dddpw}[0]{\partial^3 \phi (w)}

\newcommand{\leinsz}[0]{\lambda \phi (z)\frac{\exp \left(\mathrm{i}\sqrt{2} a \phi (z)\right)}{\mathrm{i}\sqrt{2} \alpha_0}}

\newcommand{\ba}[0]{\begin{eqnarray}}
\newcommand{\ea}[0]{\end{eqnarray}}

\def\bra{\langle}
\def\ket{\rangle}
\def\vev#1{\langle#1\rangle}
\def\vvv#1#2{\langle#1|#2\rangle}
\def\nop#1{\mbox{:$#1$:}}
\def\id{\mathbb{I}}
\def\lid{\tilde\mathbb{I}}

\begin{document}
\enlargethispage{1cm}
\maketitle
\begin{abstract}
{\small 
We examine the properties of two-dimensional conformal field theories
(CFTs) with vanishing central charge based on the extended Kac-table for
$c_{(9,6)}=0$ using a general ansatz for the stress energy tensor residing 
in a Jordan cell of rank two.

Within this setup we will derive the OPEs and two point functions 
of the stress energy tensor $T(z)$ and its logarithmic partner field $t(z)$
and illustrate this by a bosonic field realization.

We will show why our approach may be more promising than those
chosen in the literature so far, including a discussion on properties
of the augmented minimal model with vanishing central charge such
as full conformal invariance of the vacuum as a state in an irreducible 
representation, consequences on percolation from null vectors and
the structure of representations within the Kac table.

Furthermore we will present another solution to the $c \rightarrow 0$
catastrophe based on an logarithmic CFT  tensor model. As example, we consider
a tensor product of the well-known $c=-2$ logarithmic CFT with a four-fold
Ising model.

We give an overview of the possible configurations and various 
consequences on the two point functions and the OPEs of the stress energy 
tensor $T(z)$ and its
logarithmic partner field $t(z)$. We will motivate that due to the full 
conformal invariance of the vacuum at $c=0$, we have to assume
a Jordan cell for the identity since $t(z)$ is now a descendant of a new
$h=0$ field. 
}
\end{abstract}
\begin{picture}(0,0)
  \put(300,620){{\tt \ hep-th/0510096}}
  \put(300,631){{\tt BONN-TH-2005-06}}
\end{picture}

\newpage
\enlargethispage{2cm}
\tableofcontents
\newpage
\section{Introduction}
\enlargethispage{0.5cm}
During the last decade, the interest in $c=0$ conformal field theories (CFTs)
has risen considerably, 
because such theories presumably play an important role in the understanding of
percolation and other disorder problems. The problem of vanishing
central charge caused a vivid discussion on suitable approaches since
the canonical choice of ordinary minimal models seems not to be 
sufficient in its field content. There have been several attempts before,
most notably by Cardy \cite{Cardy:2001}, Kogan and Nichols \cite{Kogan:2001ku,Kogan:2002mg}
or Gurarie and Ludwig \cite{Gurarie:1999yx,Gurarie:2004ce}. 
A further clue comes from the 
deviation of the partition function from one as observed by Pearce et al.\
\cite{Pearce:2002an} in numerical studies. While the former approaches 
involve features which are not necessary for $c\neq 0$ theories, this paper 
will concentrate solely on known techniques and structures to fit $c=0$ into 
ordinary (logarithmic) CFTs. Thus, we will assume that the field content
of the $c=0$ theory can entirely be read off from an eventually extended 
Kac-table with respect to a suitable chiral symmetry algebra. Of course, the 
simplest case is the Virasoro algebra, on which we mainly focus.

In the second chapter we will start with a sketch of the problems arising
when the central charge vanishes. We know from numerical simulations, 
e.\,g.\ Pearce et al.\ \cite{Pearce:2002an}, that the partition function of 
their non-trivial $c=0$ theory is not equal to one. However, this is not
what we would expect, if the $c=0$ theory were just a plain minimal model,
since the field content as given by the Kac-table of $c_{(3,2)}=0$ consists 
only of the identity. As a direct consequence, the
field content has to be modified. For this  Cardy \cite{Cardy:2001}
gave three possible choices as explained in 2.2. From there on
we will concentrate on one of them which is based solely on the Kac table.
We will see that the Kac-table has to be extended, and the smallest
possibility for this seems to be the table for $c_{(9,6)}=0$ as discussed 
in \cite{Flohr:1996vc}.

In the third chapter we will derive the three OPEs of $T(z)$ and $t(z)$, 
starting with the assumption of two $L_0$-Jordan-cell connected $h=0$ 
fields motivated by the Kac table of the augmented minimal model
with vanishing central charge $c_{(9,6)} = 0$. Within this ansatz we find that 
only the vacuum expectation value of $t(z)t(w)$ survives. It is 
proportional to some arbitrary factor $\theta$ which equals the central 
extension of the algebra between the modes of $T(z)$ and $t(z)$. But 
contrarily to what one would expect, it is independent of the parameter 
$b$ introduced by \cite{Kogan:2001ku} and \cite{Gurarie:2004ce}, 
since it is not proportional to $\bra Tt \ket$ which itself is proportional
to the central charge $c=0$.

A direct consequence of these results will be given in 3.3 where we discuss
the impact on the OPEs of primary fields within this setup. We will show that
the metric is not invertible and thus the OPE is not well defined. This leaves
us with the question whether we should redefine the vev's in such a logarithmic 
conformal field theory (LCFT) \cite{Gurarie:1993xq} with vanishing central 
charge. For an introduction to LCFT in general, see
\cite{Flohr:2001zs,Gaberdiel:2001tr} and references therein.

In the fourth chapter we will present a bosonic free field construction
based on a vertex operator ansatz, yielding basically the same results as in our theoretical
calculation by setting $\theta = 0$. Thus we suggest that probably a
fermionic construction may provide more useful results. In general, fermionic
realizations (in contrary to bosonic ones) have the nice feature of 
a natural truncation due to their
nilpotency. Thus only terms up to $\log^2$ may arise in the OPEs of
such a theory.

Afterwards, in section 5.1 we will recall the third solution
of the $c \rightarrow 0$ catastrophe chosen by Gurarie and Ludwig
\cite{Gurarie:2004ce} or Kogan and Nichols \cite{Kogan:2001ku} which
introduces fields outside the Kac table and explicitly excludes the
case of the identity $\id$ residing in a Jordan cell. Additionally we will 
compare our approach to theirs with respect to the advantages 
and disadvantages following from the respective assumptions in 5.2.

The next chapter will elucidate features of the augmented minimal $c_{(9,6)} = 0$
model. This includes an overview on the consequences of the full
conformal invariance of the irreducible highest weight representation
generated by the identity which can be broken when regarded as
a subrepresentation of the indecomposable representation based on the
second $h=0$ field. In 6.2 we will discuss the possibility of null
states within our approach which will give us a possibility to 
subdivide augmented minimal $c=0$ models with respect to one parameter
which is the central extension of the mixed algebra between the modes
of the stress energy tensor and its logarithmic partner. This feature
will lead us to important consequences for percolation as a $c=0$
model. Additionally, we comment on the field content
of the Kac table and the structure of the representations contained
therein. Further details will appear elsewhere \cite{Holger}.

In the seventh chapter we will present a variation of the fourth loophole
added to Cardy's solutions of the $c\rightarrow 0$ catastrophe \cite{Cardy:2001} following
Kogan and Nichols \cite{Kogan:2001ku}. This tensor ansatz of an LCFT
with central charge $c_1 $ and an ordinary CFT with central charge
$c_2 = -c_1$  yields a $c=0$ theory avoiding the problems arising in the 
OPEs of primary fields. For this case we present an example of a tensorized CFT of 
symplectic fermions with a reduced fourfold Ising model in the last section.

All details of the calculations will be given in the appendix, as 
most of the results of 
the OPEs in chapters 4 and 7 were given only in parts which are sufficient
to see the important features and compare them to previous results as e.\,g.\
derived in \cite{Gurarie:2004ce}. Additionally, we will derive the algebra of
the modes of $T(z)$ and $t(z)$, $[L_n,l_m]$, to justify why we do not adopt
the result of \cite{Gurarie:2004ce}. This will include a general remark on the
mode expansion of logarithmic fields and the consequences of the requirement
of regularity.

\section{General remarks on CFTs with vanishing central charge}

\subsection[Problems at $c=0$]{Problems at \boldmath{$c=0$}}

After the introduction of conformal field theories by Belavin, Polyakov and Zamolodchikov
\cite{Belavin:1984vu} twenty years ago and the discovery of logarithmic behavior 
by Gurarie in 1993 \cite{Gurarie:1993xq} which led
to the investigation of so called logarithmic CFT, the understanding of most (L)CFTs, 
especially the minimal models characterized by the two parameters $(p,q)$ with $q,p \in 
\mathbb{N}$, $c_{(p,q)}=1-6\frac{(p-q)^2}{pq}$ improved continually.
As to CFTs whose field content can not be described solely by the Kac table,
i.\,e.\ a non-trivial $c_{(3,2)}=0$ model, this is not the case. There is still a 
controversial discussion going on about different approaches to (L)CFTs
with vanishing central charge which we will try to elucidate in this paper.

For $c=0$ as an ordinary minimal model, we have $(p,q)=(3,2)$ and thus a Kac table
which consists only of one field,
the identity. Keeping the vanishing of the central charge in mind, we know
that $L_n |0 \ket = 0$ for all $n \in \mathbb{Z}$ and thus the theory is trivial.
But from concrete models, e.\,g.\ by Pearce and Rittenberg \cite{Pearce:2002an}, we
know that the partition function differs from one and therefore there have to be
more fields involved. More concretely, they identify an $h=1/3$ primary field.

A similar problem occurred in the study of the $c_{(p,1)}$ models whose Kac-tables
a priori are empty. Following the procedure which is usually applied to this kind of
minimal model, i.\,e.\ including the operators on the boundary of the conformal grid into
the theory, we get a non trivial $c=0$ CFT. Additionally it can be shown that they 
generate indecomposable representations which leads to logarithms in the OPEs of
some of these fields \cite{gaberdiel-1996-477}. 
The main advantage of this procedure
is that we maintain the properties of all finite Kac-table based CFTs, e.\,g.\ the existence of
an infinite set of null vectors, thus a rather small field content and the possibility of additional
symmetries. Remarkably, up to now in all known logarithmic CFTs, i.\,e.\ the $c_{(p,1)}$ 
models, the identity has a logarithmic partner. Taking the well known formula
\begin{equation}
h_{r,s}(c_{(p,q)}) = \frac{(pr-qs)^2-(p-q)^2}{4 pq}
\end{equation}
for $q=1$ and $1 \leq s <3p$, $1 \leq r < 3q$, i.\,e.\ the weights of the operators on 
the boundary of the conformal grid and those needed for closure under fusion, 
we always have at least two solutions for $h=0$: $s_{\pm} = rp \pm (p-1)$. 
In a logarithmic theory, these cannot be identified with each other. Thus 
taking a similar approach to construct a non-trivial $c=0$ LCFT, we would expect
it to contain a degenerate vacuum as well.

There exists a variety of proposals on how to approach $c=0$. Apart from the suggestions
of other LCFTs as discussed above, Cardy \cite{Cardy:2001} tried a general replica 
ansatz in order to find a loophole to the divergences arising in the OPE of primary
fields at $c=0$.

For any conformal field theory (for the time being we will restrict ourselves to
non degenerate vacua and the holomorphic parts), we can write down the OPE of 
a primary scalar field $\phi(z)$ with conformal weight $h$, 
\begin{equation}\label{eq:OPE}
\phi_h(z) \phi_h^\dagger(0) \sim \frac{C^{\id}_{\Phi\Phi}}{z^{2h}}\left( 1 + \frac{2h}{c} z^2 T(0) + \ldots
\right) + \ldots \, ,
\end{equation}
with $C^{\id}_{\Phi\Phi}$ being the coefficient of the three point
function usually normalized to $1$ or $\frac{c}{h}$ for $h\neq 0$.
For the ordinary minimal model $c_{(3,2)}=0$ the expression is not problematic
since the only possible choice for $\phi$ is $\id$ and thus $h=0$. Although, 
if we seek to describe a model as found by \cite{Pearce:2002an}, we have to
assume additional fields to the identity for which the division by the central charge is not
well defined.
\subsection[Suggestions how to treat $c=0$ properly]{Suggestions how to treat \boldmath{$c=0$} properly}\label{cardy}
According to Cardy \cite{Cardy:2001}, there are basically three ways out of the
problem as the central charge approaches zero in the OPE of primary fields as given in (\ref{eq:OPE}).
\begin{itemize}
\item[(I)] \quad $(h,\bar{h}) \rightarrow 0$ as $c \rightarrow 0$.
\item[(II)] \quad $C^{\id}_{\Phi\Phi} \rightarrow 0$  as $c \rightarrow 0$.
\item[(III)] \quad Other operators arise in the OPE, canceling the divergencies.
\end{itemize}
Thus the first case can be applied to the ordinary minimal model with the Kac-table
\begin{equation}
c_{(2,3)} = 0 \quad : \quad
\begin{array}{|c|c|}
\hline
\rule[-.3em]{0cm}{1.3em} 
0&0\\
\hline
\end{array}\, .
\end{equation} 
The second case has to be taken if we restrict ourselves to the extended Kac-table 
as for the $c_{(p,1)}$-models. In this case we have to normalize our three-point functions
to $\frac{c}{h_{\phi}}$ and thus the condition $C^{\id}_{\Phi\Phi} \rightarrow 0$ as 
$c \rightarrow 0$ is satisfied trivially and the OPE 
\begin{equation}
\phi_h(z) \phi_h^\dagger(0) \sim \frac{c}{hz^{2h}} + \frac{2}{z^{2h-2}} T(0) + \ldots
\end{equation}
stays regular. 
As discussed above, we expect the identity to
have a partner field in these theories, and thus we have to modify the OPE of primary 
fields. This will be done in the third chapter.

The third case has been chosen by Kogan and Nichols \cite{Kogan:2001ku} as well as by 
Gurarie and Ludwig \cite{Gurarie:2004ce}. It includes a new concept of LCFTs which is
structurally different from that of $c_{(p,1)}$ models. Introducing the limiting 
procedure to get a logarithmic partner of the stress energy tensor, fields outside of the
Kac-table arise in the OPE of primary fields. We do not have any knowledge about their
behavior in OPEs among themselves and thus there is no a priori limit on the number of
fields available in the emerging CFT. Nothing of what is known for Kac-table based
models as null states, symmetries or representation properties can be assumed to be
extended to this kind of $c=0$ theory. Furthermore, this approach introduces fields
that have no known direct physical meaning at all since in all known applications for
$c=0$ the critical exponents of physical quantities are expected to be out of 
the Kac table. Thus it seems more natural to stay within this framework, or, 
more precisely, include the operators on the boundary of the Kac-table.
For example, in the 
work of Pearce et al. \cite{Pearce:2002an}, the representation belonging to 
$h_{(1,3)}=1/3$ is expected to play an important role.

%
These are the crucial points where we do not agree with the approach of 
\cite{Cardy:2001} who suggested that the logarithmic partners of Kac-operators
always reside outside of the Kac-table. In contrary, we favor the ansatz of
sticking to the restrictive structure of an augmented minimal
model following e.\,g.\ \cite{Flohr:1996vc} taking the Kac-table of $c_{(9,6)}$ as
a basis to describe a $c=0$ LCFT.
%
%
\section{OPEs in the augmented minimal model }
\subsection{The standard assumptions for two-point functions}
In the following we will show how our approach differs from the usual 
constructions. In contrary to our ansatz it is usually assumed that the Jordan
cell exists on the $h=2$ level and \textit{not} on the identity level as well. 
But in our opinion any theory with arbitrary central charge $c\neq 0$, if
extended to a logarithmic CFT, has to possess a global Jordan cell structure. 
In standard LCFT, primary fields and their
logarithmic partners form Jordan cells with respect to $L_0$. 
The identity always resides in such a Jordan cell
and, particularly, there can not be a Jordan cell structure at the second
level without having this structure in the vacuum sector, although this is 
the basis for the calculations of \cite{Gurarie:2004ce,Kogan:2001ku,Nagi:2005sb,Nagi:2005cm}. 
At least we do not know of any LCFT
whose identity does not reside in an indecomposable representation.

A consequence of this consideration is the following crucial property of vacuum 
expectation values (see e.g. \cite{Rasmussen:2005ta} or \cite{Flohr:2001tj} for an elaborate treatment), where $\lid(z)$ denotes the logarithmic partner of
the identity $\id$:
\begin{eqnarray*}
  \vvv{0}{0} = \bra 0|\id|0\ket &=& 0\,,\\
  \vvv{\tilde 0}{0} = \vvv{0}{\tilde 0} = \bra 0|\lid(z)|0\ket &=& 1\,,\\
  \bra 0|\lid(z)\lid(w)|0\ket &=& -2\log(z-w)\,.
\end{eqnarray*}
If we keep this in mind, it is clear that any stress tensor within a LCFT
will have a vanishing two-point function, where $n$-point functions are
understood as usual, as vacuum expectation values of $n$ field insertions.
This must be so, because the central term comes with the identity $\id$ and
not its logarithmic partner, i.\,e.\
\begin{equation}\label{eq:TT}
  T(z)T(w) = \frac{c/2}{(z-w)^4}\id + \frac{2}{(z-w)^2}T(w) 
           + \frac{1}{(z-w)^{}}\partial T(w)\,.
\end{equation}

Suppose now that the stress energy tensor $T(z)$ has a logarithmic partner,
which we call $t(z)$ to match with the convention in the literature. In fact,
there is a rather simple example of how to construct such a partner, $t(z) = \nop{\lid T}(z)$.
Standard considerations in LCFT now imply the following behavior in an
OPE with the stress energy tensor, revealing the Jordan cell structure:
\small
\begin{equation}\label{eq:Tt}
  T(z)t(w) = \frac{c/2}{(z-w)^4}\lid + \frac{\mu}{(z-w)^4}\id
           + \frac{2}{(z-w)^2}t(w) + \frac{\lambda}{(z-w)^2}T(w)
           + \frac{1}{(z-w)^{1}}\partial t(w)\,.
\end{equation}
\normalsize
Note that the central charge $c$ now appears together with the logarithmic
partner $\lid$ of the identity, and that a new central term $\mu$ may appear
with the identity $\id$. Also, the constant $\lambda$ with which the original
stress energy tensor appears, can be any non-zero number depending on the
normalization of the off-diagonal part in the Jordan cell. It is conventional
to put $\lambda=1$. As standard logarithmic pair, $\{T,t\}$ must 
obey the following two-point functions
\begin{eqnarray}
  \label{eq:vTT}\vev{T(z)T(w)} &=& 0\,,\\
  \label{eq:vTt}\vev{T(z)t(w)} &=& \frac{b}{(z-w)^4}\,,\\
  \label{eq:vtt}\vev{t(z)t(w)} &=& \frac{1}{(z-w)^4}\left(\theta 
     - 2b\log(z-w)\right)\,.
\end{eqnarray}
We once more emphasize that this should be true in any LCFT, independent of
the value of the central charge $c$.

\newpage
\subsection{The generic form of the OPE} 
We will now start with 
the generic form of the OPE for a pair of two fields of the same weight either
being the primary field or its logarithmic
partner, respectively, following \cite{Flohr:2001tj,Flohr:2003tc}. 
%
In general, it is given by
\begin{equation}\label{eq:OPELCFT}
\phi_{h_i}(z) \phi_{h_j}(0) = \sum_k C_{ij}^k (z-2)^{h_k}\left(\phi_{h_k}
+ \sum_{\lbrace n \rbrace} \beta_{ij}^{k,\lbrace n \rbrace} (z-w)^{|\lbrace n \rbrace|}
\phi_{h_k}^{(-\lbrace n \rbrace)}(w) \right),
\end{equation}
where the coefficients $\beta_{ij}^{k,\lbrace n \rbrace}$ of the descendant 
contributions, 
\begin{equation}
\phi_{h_k}^{(-\lbrace n \rbrace)} = L_{(-\lbrace n \rbrace)} \phi_{h_k}
= L_{-n_1} L_{-n_2} \ldots L_{-n_l} \phi_{h_k}\, ,
\end{equation}
are fixed by conformal covariance.

The structure ``constants'' $C_{ij}^k$ (which in an LCFT can no longer referred to
as constants since they merely become functions containing logarithms) can be derived 
through the two- and three-point
functions, i.\,e.\ $C_{ij}^k = C_{ijl} D^{lk}$, with
\begin{eqnarray}
D_{ij} &=& \bra \phi_{h_i} (\infty) \phi_{h_j}(0) \ket \propto \delta_{h_i,h_j} \, ,\\
C_{ijk} &=& \bra \phi_{h_i} (\infty) \phi_{h_j} (1) \phi_{h_k}(0) \ket \, .
\end{eqnarray}
Note that in our case of an LCFT the metric is no longer diagonal but for 
$h \equiv h_i = h_j$ looks like
\begin{equation}
D_{(i,j)} = \left(\begin{array}{cc} 0 & D_{\Phi\Phi}^{(0)} \\
D_{\Phi\Phi}^{(0)} & D_{\Phi\Phi}^{(1)} - 2 D_{\Phi\Phi}^{(0)} \log (z-w)
\end{array}\right) (z-w)^{-2h}
\end{equation}
in the notation following below. 

In fact, within the most general ansatz of a rank two Jordan cell, 
we have for the two-point functions
\small
\begin{eqnarray}\label{eq:notation}
  \vev{\Phi(z)\Phi(w)} &=& 0\,,\\
  \vev{\Phi(z)\tilde\Phi(w)} = \vev{\tilde\Phi(z)\Phi(w)} 
    &=& D_{\Phi\Phi}^{(0)}(z-w)^{-2h}\,,\\
  \vev{\tilde\Phi(z)\tilde\Phi(w)} 
    &=& \left(D_{\Phi\Phi}^{(1)} - 2\log(z-w)D_{\Phi\Phi}^{(0)}\right)
    (z-w)^{-2h}\,,
\end{eqnarray}
\normalsize
where $D_{\Phi\Phi}^{(0)}=D_{\Phi\tilde\Phi}=D_{\tilde\Phi\Phi}$ and 
$D_{\Phi\Phi}^{(1)}=D_{\tilde\Phi\tilde\Phi}$.
For the three-point functions involving the pair $\{\Phi,\tilde\Phi\}$, 
omitting the explicit dependence on the three points of insertion
$z_1,z_2,z_3$, we have
\begin{eqnarray*}
  \vev{TT\Phi} &=& 0\,,\\
  \vev{TT\tilde\Phi} &=&
  C_{TT\Phi}^{(0)}z_{12}^{h-4}z_{13}^{-h}z_{23}^{-h}\,,\\
                     &=& \vev{tT\Phi} = \vev{Tt\Phi}\\  
  \vev{tt\Phi} &=& 
    \left(C_{TT\Phi}^{(1)}-2\log(z_{12})C_{TT\Phi}^{(0)}\right)
                 z_{12}^{h-4}z_{13}^{-h}z_{23}^{-h}\,,\\
  \vev{tT\tilde\Phi} &=& 
    \left(C_{TT\Phi}^{(1)}-2\log(z_{13})C_{TT\Phi}^{(0)}\right)
                 z_{12}^{h-4}z_{13}^{-h}z_{23}^{-h}\,,\\
  \vev{Tt\tilde\Phi} &=& 
    \left(C_{TT\Phi}^{(1)}-2\log(z_{23})C_{TT\Phi}^{(0)}\right)
                 z_{12}^{h-4}z_{13}^{-h}z_{23}^{-h}\,,\\
  \vev{tt\tilde\Phi} &=& \left(C_{TT\Phi}^{(2)} 
    - C_{TT\Phi}^{(1)}(\log(z_{12})+\log(z_{13})+\log(z_{23}))\right.\\
    & & \left.\vphantom{C_{TT\Phi}^{(0)}}
      - C_{TT\Phi}^{(0)}(\log^2(z_{12})+\log^2(z_{13})+\log^2(z_{23})
      - 2\log(z_{12})\log(z_{13})\right.\\ 
    & & \left.\vphantom{C_{TT\Phi}^{(0)}} 
      - 2\log(z_{12})\log(z_{23})
      - 2\log(z_{13})\log(z_{23}))\right)
                 z_{12}^{h-4}z_{13}^{-h}z_{23}^{-h}\,.
\end{eqnarray*}
Note that the constants in front of the logarithms are always given in terms
of the constant of previous correlators with less insertions of
logarithmic partner fields. Moreover, the constants depend only on the
total sum of logarithmic insertions. We are interested in two particular
contributions to the OPE, namely the pair $\{\id,\lid\}$ and the pair
$\{T,t\}$ for $\{\Phi,\tilde\Phi\}$, respectively. But let us first write down the
generic form of the OPE channel with conformal weight $h$, expressed in
the structure constants of the two- and three-point functions:
\small
\begin{eqnarray*}
  T(z)T(0) &=& z^{h-4}\frac{C_{TT\Phi}^{(0)}}{D_{\Phi\Phi}^{(0)}}\Phi(0)\,+ \ldots \quad,\\
  T(z)t(0) &=& z^{h-4}\left(
            \frac{C_{TT\Phi}^{(0)}}{D_{\Phi\Phi}^{(0)}}\tilde\Phi(0)
            + \frac{C_{TT\Phi}^{(1)}D_{\Phi\Phi}^{(0)} 
                    - C_{TT\Phi}^{(0)}D_{\Phi\Phi}^{(1)} 
            }{(D_{\Phi\Phi}^{(0)})^2}\Phi(0)\right)\,+ \ldots \quad,\\
  t(z)t(0) &=& z^{h-4}\left[\left(
            \frac{C_{TT\Phi}^{(1)}}{D_{\Phi\Phi}^{(0)}}
            - 2\log(z)\frac{C_{TT\Phi}^{(0)}}{D_{\Phi\Phi}^{(0)}}\right)
            \tilde\Phi(0)+\left(\frac{C_{TT\Phi}^{(2)}D_{\Phi\Phi}^{(0)} 
                                     - C_{TT\Phi}^{(1)}D_{\Phi\Phi}^{(1)} 
                                    }{(D_{\Phi\Phi}^{(0)})^2}\right.\right.\\
           &\quad& \left.\left. -\log(z)\frac{C_{TT\Phi}^{(1)}D_{\Phi\Phi}^{(0)} 
                    - 2C_{TT\Phi}^{(ß0)}D_{\Phi\Phi}^{(1)} 
            }{(D_{\Phi\Phi}^{(0)})^2}
            -\log^2(z)\frac{C_{TT\Phi}^{(0)}}{D_{\Phi\Phi}^{(0)}}
            \right)\Phi(0)
            \right]+ \ldots \quad.
\end{eqnarray*}
\normalsize
We are now in the position to fix most of the constants. 
The typical normalization for the identity channel is $D_{\id\id}^{(0)}=1$,
whereas $D_{\id\id}^{(1)}=d$ is left undetermined. But from our ansatz
(\ref{eq:vTT}) - (\ref{eq:vtt}), we
also know the normalization for the channel of the stress energy tensor,
namely $D_{TT}^{(0)}=b$ and $D_{TT}^{(1)}=\theta$. Furthermore, we know how
the OPE of the stress energy tensor with itself and with its logarithmic
partner must look like, see eqs.\ (\ref{eq:TT}) and (\ref{eq:Tt}). This
allows us, by comparing coefficients, to fix further constants, namely
$C_{TTT}^{(0)}=2b$, $C_{TTT}^{(1)}=\lambda b+2\theta$ and
$C_{TT\id}^{(0)}=c/2$, $C_{TT\id}^{(1)}=\mu+cd/2$. These choices are all
natural and then exactly reproduce the OPEs $TT$ and $Tt$ as given in 
eqs.\ (\ref{eq:TT}) and (\ref{eq:Tt}). 

Before we continue, we have to address one issue of consistency. So far,
we have tried to choose the normalization of the two-point functions
of the stress energy tensor and its partner independently of the central
charge of the theory. However, these two-point functions do not change,
if we insert the identity as third field. Therefore, the structure
constants must obey the relations $D_{Tt}=C_{Tt\id}=C_{TT\lid}$ and
$D_{tt}=C_{tt\id}=C_{Tt\lid}$. Hence, $D_{TT}^{(0)}=C_{TT\id}^{(0)}$ and
thus $b=c/2$. Furthermore, $D_{TT}^{(1)}=C_{TT\id}^{(1)}$ and thus
$\mu=\theta-cd/2$. As a consequence, the only remaining free parameters
are the central charge $c$ and the normalizations $d$ and $\theta$ in the
tow-point functions $\vev{\id\lid}$ and $\vev{\lid\lid}$.
Plugging these choices into the
remaining OPE $tt$ as given above, yields the following structure:
\begin{eqnarray}\label{eq:generell}
  t(z)t(0) &=& 
           z^{-4}\left(\theta-\log(z)c\right)\lid(0)\nonumber\\
         &+& z^{-4}\left(C_{TT\id}^{(2)}-\theta d +
                 \log(z)(cd-\theta)-\log^2(z)\frac{c}{2}\right)\id\nonumber\\
         &+& z^{-2}\left(1+\frac{4\theta}{c} - 4\log(z)\right)t(0)\nonumber\\
         &+& z^{-2}\left(2\frac{C_{TTT}^{(2)}-\theta}{c}-\frac{4\theta^2}{c^2}
                 - \log(z)(1-\frac{4\theta}{c})-2\log^2(z)
                 \right)T(0)\nonumber\\
		&\quad& + \ldots \quad.
\end{eqnarray}
The choice $\theta=0$ reproduces the result of Gurarie and Ludwig. In our
approach, we only see the primary fields and the logarithmic partners, but
none of the descendants. This is, why our formula misses the canonical
terms proportional to $\partial t$, $\partial T$ and $\partial\lid$.

The result is also very similar to the one derived in \cite{Kogan:2001ku} 
(but only for $\theta = 0$, too). Allowing for the terms containing $\lid$
to differ, we see that the two characteristic parameters can be described
by the central charge and $C^{(2)}_{TTT}$. But $\theta = 0$ implies a
vanishing two-point-function for $\bra T t \ket$ and $\bra tt \ket$, at 
least for our ansatz of a Jordan cell structured identity sector with 
$b = \frac{c}{2} = 0$. Thus if our ansatz is correct we are not allowed to
throw these parameters away. 

There is one very important caveat: The most singular term of the OPE
stems from the pair $\{\id,\lid\}$, and starts with $z^{-4}$. The higher level
descendants of these fields will influence the terms of the OPE of order
$z^{-2}$. This is particularly important, because these descendants
are largely given in terms of $T$ and $t$. Indeed, we have
$L_{-2}\id = T$ and $L_{-2}\lid = t + \lambda T$. 

Most of the results concerning the stress energy tensor have also been derived
in \cite{Moghimi-Araghi:2000qn}, especially $b = \frac{c}{2}$ and the 
normalization of the Jordan cell of the stress energy tensor. 

\newpage
\subsection[Consequences on the $c \rightarrow 0$ catastrophe]{Consequences on the \boldmath{$c \rightarrow 0$} catastrophe}
The impact on the OPE of primary fields of the results derived in the previous
section is immense. 

Therefore let us recall what we know about the general form of the OPE in
equations (\ref{eq:OPELCFT}ff).
Inserting the fact that for our ansatz (and $h=0$) we have 
$D_{\Phi\Phi}^{(0)} \propto \bra \frac{c}{2} \lid (w) + \mu \id \ket =0$, 
we run into a problem inverting the matrix of two-point functions 
which is needed to raise indices, since $D^{ij} =(D_{ij})^{-1}$.
Hence the OPE of two primary fields in a $c=0$ theory with a Jordan
cell structure on the $h=0$ level and $T(z),t(z)$ being descendants of 
the $h=0$ fields, remains ill defined. The only loophole to this is to
define the normalization of the three-point functions to $\frac{c}{h_\phi}$.
Thus for $h \neq 0$, the metric would be invertible again following the
suggestion of Cardy \cite{Cardy:2001}.

Since for $h=0$ the problem is not solved yet, this brings up the question 
whether for $c=0$ we can still stick to our usual 
definition of vacuum expectation values or if we should simply redefine the
vev to be proportional to 
\begin{equation}
\bra \cdot \ket : = \bra 0 | \cdot | \tilde{0} \ket + \bra \tilde{0} | \cdot | 0 \ket,
\end{equation}
leaving us with the problem of how expressions like the vev of 
the OPE of $t(z)t(w)$ may be dealt with. A motivation for this behavior
may be found in \cite{Gurarie:1997dw} where the vanishing of the fermionic path 
integral in the $c=-2$ LCFT is discussed.
\section{A bosonic free field construction}
\subsection{Ansatz}
To illustrate the results obtained from the most singular term of the OPE 
by global conformal invariance, we take a free field construction with 
arbitrary central charge for the stress-energy tensor and its logarithmic 
partner field:
\begin{eqnarray}
T(z) &=& -\frac{1}{2} \nop{\dpz \dpz } + \cons \nop{\ddpz } \, ,\\
t(z)&=& \nop{\leinsz T(z)}\, .
\end{eqnarray}
For the logarithmic partner of the identity we chose a vertex operator ansatz
with conformal weight $h(a) = a^2 - 2 a \alpha_0 = 0$ which means that we
have two possible weights for the Vertex operator, $a = 0$ for the true
identity and $a = 2 \alpha_0$ for the second. Thus we expect another
vertex operator to appear in the OPE behaving like the identity in correlators.
This non-trivial ansatz for the logarithmic partner of the identity is 
necessary to reproduce the standard form of the OPEs typical for LCFT.
Thus we define with $a=2\alpha_0$:
\begin{eqnarray}
\lid (z) &=& \leinsz \, , \\
\id' (z) &=& \exp (\mathrm{i} \sqrt{2} a \pz )\ \equiv\id\, .
\end{eqnarray}
Similar considerations can be found within the Coulomb gas formalism
used in \cite{Kogan:1997fd}.
\subsection{The non logarithmic OPEs}
Of course, the OPE of the stress energy tensor with itself is as usual,
\begin{equation}
T(z)T(w) \sim \frac{\frac{c}{2}}{(z-w)^4}+\frac{2T(w)}{(z-w)^2} + \frac{\del T(w)}{(z-w)}\, ,
\end{equation}
thus, due to the vanishing vev of $\id$, we have
\begin{equation}
\bra T(z)T(w) \ket = 0.
\end{equation}
The first different result appears in the correlator of the stress energy tensor 
with its logarithmic partner:
\begin{equation}
T(z)t(w) \sim \frac{\frac{c}{2} \lid(w)}{(z-w)^4} + \frac{2 t(w)+\lambda T(w)}{(z-w)^2}+ \frac{\del t(w)}{z-w} \, ,
\end{equation}
where $\lambda$ depends on the normalization of the off-diagonal entries of the 
Jordan cell between $T(z)$ and $t(z)$, i.\,e.\ $L_0 t(z) = 2 t(z) + \lambda T(z)$. 
These two are exactly what we calculated before. Since the vev of $\lid$ does not vanish in an LCFT, 
we are left with a vacuum expectation value of
\begin{equation}
\bra T(z)t(w) \ket = \frac{\frac{c}{2}}{(z-w)^4} = 0.
\end{equation}
However, for $c=0$ this vanishes, too.

Note that there is a subtlety here. This ansatz can not be directly compared to the general
formula, especially not for $t(z)t(w)$ as in (\ref{eq:generell}) since the propagator is not of the standard
form. We do not only have the identity $\id$ and its logarithmic partner field
$\lid$ but in addition also a field that is conjugated to the identity, 
$\id' = \exp (\mathrm{i}\sqrt{2} a \phi)$ with $a = 2 \alpha_0$. Thus contributions
by this field have to be taken into account, too. This leads to different
prefactors and changes in signs. Additionally, it is not surprising that we are
not able to get the coefficients of $T(w)$ on the rhs of $T(z)t(w)$ and those
of the most singular terms in the OPE of $t(z)t(w)$ to overlap. This is simply
due to the fact that the normalization of the Jordan cell of the stress energy
tensor is already fixed by that of the identity. This
is the reason why some factors appear twice as often as expected when compared to the OPE derived
by Gurarie and Ludwig \cite{Gurarie:2004ce} or Kogan and Nichols \cite{Kogan:2001ku}.
\subsection{The logarithmic OPE}
The OPE of $t(z)$ with itself is rather more complicated as one would expect from
the derivation from its most singular term due to the appearance of descendants of 
$\id$ and $\lid$ as already mentioned above. Another point to bear in mind are the 
divergences of lowest order, $\log (z-w) (z-w)^0$, which have been omitted in the 
literature before.

To keep things as simple as possible, we will just state our result for the case that
$4-2a^2 >0$ and omit the terms that are dispensable for the comparison with the
results of \cite{Gurarie:2004ce} and our general calculation (\ref{eq:generell})
 which means that we will restrict ourselves
to the contributions of $T, t, \id$ and $\lid$ up to first order without logarithmic
divergences or composite fields. The full results will be given in the appendix.
Choosing $\lambda = \frac{1}{2}$ and for $c=0$, $\alpha_0^2 = \frac{1}{24}$ we get:
\small
\begin{eqnarray*}
&&t(z)t(w) \\
&\sim& \left(- 2\lzw\lid (w) + \frac{1}{2c} + \log^2 (z-w) + 3 \lzw \right) \frac{\frac{c}{2}}{(z-w)^4}\\
&\quad& +\frac{(1-4\lzw) t(w) + \left(3 \lzw + 2 \log^2 (z-w)\right)T(w)}{(z-w)^2} \\
&\quad& + \frac{(1-4 \lzw) \del t(w) + \left(3\lzw+ 2\log^2 (z-w)\right)\del T(w)}{2(z-w)}\, .
\end{eqnarray*}
\normalsize
Of course, the first line vanishes for $c=0$ (since $\bra 0 | \id | 0 \ket = 0$, too) 
but since these terms may be
interesting for the reader, we stated them in spite of that.
Additionally we should mention that it is obviously possible to chose a $\lid$ which is
quasi-primary since its OPE with the stress energy tensor looks like
\begin{equation}
T(z) \lid(w) \sim \frac{\lambda \id}{(z-w)^2} + \frac{\partial_w \lid(w)}{(z-w)}\, .
\end{equation}
%
Recapitulating, we have shown that as in \cite{Gurarie:2004ce}, we have $\theta =0$, 
which means that we would have vanishing vacuum expectation values for $c=0$. 
These results suggest that we can not take a naive free field construction to describe the 
problem, since all vev's vanish. In order to get the necessary central
extensions, more sophisticated constructions should be considered, such 
as several free fields or deformations of the stress energy tensor similar
to the ones introduced in \cite{Fjelstad:2002ei}.
In contrary to their assumption, we found a bunch of 
more complicated fields than $t(w)$, $T(w)$ or their descendants though there are no 
other primaries involved. 

Therefore we tried to find a way out by searching for $\sum_{\lbrace p,q \rbrace} c_{p,q}=0$
to construct a $c=0$ theory out of tensorized minimal models to 
get a non trivial CFT with vanishing central charge in chapter seven.

Another possibility would be to find a fermionic theory with vanishing central charge.
\section{Discussion of the two LCFT approaches}
As already stated in section \ref{cardy}, we have a third possible loophole
to avoid the $c \rightarrow 0$ catastrophe in the OPEs of primary fields. In the
following we will explain why the ansatz we chose, i.\,e.\ the one based on the augmented
minimal model, may be a more natural solution.

Hence we will give a brief overview on the ansatz of Kogan and Nichols
\cite{Kogan:2001ku} followed by our comments on their approach. Additionally
we will state some facts about the $c=0$ case including implications for
percolation and a discussion of current research on augmented $c_{(p,q)}$ models
with $q > 1$ which have not been treated in the literature so far, focusing
on $p=3, q=2$.
\subsection{The replica approach to vanishing central charge}
Following the replica approach of \cite{Cardy:2001}, Kogan and Nichols
\cite{Kogan:2001ku} introduced another field $\tilde{T}$ with dimension 
$h=2 + \alpha (c)$ which satisfies $ \alpha(c) \rightarrow 0$ for $c \rightarrow 0$ 
being normalized to
\begin{equation}
\bra \tilde{T}(z) \tilde{T} (0) \ket = \frac{1}{c} \frac{B(c)}{z^{4+2\alpha(c)}}
\end{equation}
with $B(c) = -\frac{h^2}{2}+B_1c + \ldots$\quad. Thus for $c \rightarrow 0$ this 
expression diverges. 

Then, after a small $c$ expansion, the OPE of our primary field looks like
\small
\begin{eqnarray*}
\phi_h(z) \phi_h^\dagger(0) &\sim& \frac{1}{z^{2h}}\left( 1 + \frac{2h}{c} z^2 T(0) +
2 z^{2 + \alpha (c)}\tilde{T}(0) + \ldots \right) + \ldots\\
&\sim& \frac{1}{z^{2h}} + \frac{1}{z^{2h}} z^{2 + \alpha (c)}
\left( \frac{2h}{c} \left(1-\alpha(c) \log (z) \right) T(0) 
+ 2 \tilde{T}(0) + \ldots \right) + \ldots \quad ,\\
\end{eqnarray*}
\normalsize
which again is well defined, if $\mu$ which is given by
\begin{equation}
\mu^{-1} \equiv \lim\limits_{c \rightarrow 0} -\frac{2\alpha(c)}{c} = -2\alpha '(c)\, ,
\end{equation}
is not equal to zero. 

The logarithmic partner field can now be defined by
\begin{equation}
\frac{h}{\mu} t = \frac{2h}{c} T + 2\tilde{T}\, ,
\end{equation}
satisfying
\begin{equation}
L_0 T = 2T \quad L_0 t = 2t + T.
\end{equation}
This means that $t(z)$ is a field of the same conformal weight living
in a Jordan cell due to $L_0$ being non-diagonalizable. Thus the OPE
becomes 
\begin{equation}
\phi_h(z) \phi_h^\dagger(0) \sim \frac{1}{z^{2h}}
\left(1 +  \frac{h}{\mu} \left(t(0)-\log (z) T(0)\right) + \ldots \right) + \ldots\quad ,
\end{equation}
which yields the following vevs after redefining $t \rightarrow t + \gamma T$
with a suitable choice of $\gamma$
\begin{eqnarray}
\bra T(z)T(0) \ket &\sim& 0 \, ,\\
\bra T(z)t(0) \ket &\sim& \frac{b}{z^4} \, , \label{eq:vevTtNichols}\\
\bra t(z)t(0) \ket &\sim& \frac{-2b \log z}{z^4} \, .
\end{eqnarray}
This result can only be obtained by assuming that we are dealing with non-
degenerate vacua, which means, that the vacuum expectation value of the identity
operator does not vanish and we have only one $h=0$ field contributing to the OPE. 
It is based on the following algebra between the modes of
$T(z)$ and $t(z)$:
\begin{equation}
[L_n, t(z)] = z^n \left\lbrace\left( z \frac{\mathrm{d}}{\mathrm{d}z} 
            + z (n+1) \right) t(z) 
            + (n+1)T(z)\right\rbrace + \frac{\mu \id }{6}n(n^2+1)z^{n-2} \, ,
\end{equation}
where for their ansatz, we have $\mu = b$.

Thus, since $\bra \id \ket \neq 0$, the most singular part of the OPE
yields the vacuum expectation value as stated in (\ref{eq:vevTtNichols}).
This is only true if we assume $L_{-2}|0\ket = T(0)|0\ket$ not to be
zero by construction since the action of the conformal generators on the
vacuum vanish in a $c=0$ CFT but to be some kind of generalized null 
state on which $l_{-2}|0\ket = t(0)|0\ket$ is non orthogonal. To keep
this assumption it is crucial not to have a logarithmic partner of the
identity and thus $\bra \id \ket \neq 0$.
\subsection{Comments on the replica approach}
However, we have a few comments on this treatment of $c=0$ LCFTs. 

A rather small one is about the fact that
for all $c\neq 0$ $\mu$ may be set to zero by a redefinition
$l_m \rightarrow l_m - \frac{2\mu}{c} L_m$. Hence it is not obvious how this 
limit may equal the value of $\mu^{-1}$ for $c=0$ due to the discontinuity
of being free to choose $\mu=0$ for $c\neq 0$ but staying with fixed $\mu$ 
for $c=0$. Furthermore, there is no physical quantity known to correspond
to this arbitrary parameter $\mu$, thus it is rather awkward that it may
show up with such a significant role in our (L)CFT.

Additionally, we doubt that it is possible to choose the vacuum as a
``stand alone'' irreducible representation not contained in an indecomposable
one based on a second $h=0$ state, called $\lid$. We will see that in a Kac
table based (L)CFT ansatz for $c=0$ we have indeed three $h=0$ fields
which seem to belong to rank three Jordan cell structures whose details
are not yet clarified. Within the setup we chose based on the knowledge
that there is more than one $h=0$ field ($\id $) which is most probably connected
to another $h=0$ field ($\lid$) via $L_0$ and thus an at least rank two
Jordan cell structure, we know that the two point function of $\bra Tt \ket$
has to vanish and thus the parameter $b$, too, as we will see in the following.

For Kogan and Nichols \cite{Kogan:2001ku}, the term proportional to the identity
in the central extension of the algebra between the Laurent modes of $t(z)$ and $T(z)$, 
$\mu$, is the same as the proportionality factor of $\bra Tt \ket$.
In our calculations, however, they are different since we assume a Jordan cell
on the identity level, yielding
\begin{equation}
[L_n, t(z)] = z^n \left\lbrace\left( z \frac{\mathrm{d}}{ \mathrm{dz}} + z (n+1) \right) t(z) + 
				(n+1)T(z)\right\rbrace + \frac{\mu \id + \frac{c}{2}\lid}{6}n(n^2+1)z^{n-2}\, ,
\end{equation}
or, equivalently $T(z) t(0) \sim (\mu \id + b \lid)z^{-4} + \ldots$\quad. 
Thus $b = \frac{c}{2}$ in our case and $\bra Tt \ket$ has to vanish, too.
A priori, as already discussed above, there is no constraint on the choice 
of $\mu$. Following Gurarie and Ludwig
\cite{Gurarie:2004ce}, we will show how various values of $\mu$ affect the theory in 
the next section.

Thus we have motivated that in a $c_{(9,6)}=0$ (L)CFT, there is no level two state
which is non orthogonal to $T(0)|0\ket$, especially not $t(0)|0 \ket$ since the
two point function has to vanish. Thus in this setup, we can keep the full
(and not only global) conformal invariance of the vacuum. This may lead to 
consequences on the construction of the stress energy tensor.

\section{The \boldmath{$c_{(9,6)}=0$} augmented minimal model}
\subsection{Consequences of full conformal invariance}
If we restrict ourselves to the case of not having a Jordan cell structure at the
($h=0$)-level, we encounter 
the fact that any two-point function involving $T$ has
to vanish. This follows directly from the behavior of the identity sector in a ($c=0$)-theory
unless we introduce a non orthogonal state to $L_{-2} |0\ket$.
We know that by global conformal invariance and the highest weight condition, we have
$L_n \id = L_n | 0 \ket = 0$ for all $n \geq -1$. In the following, let $n$ be
$>0$. Starting with a vanishing central charge and $h=0$, we know
\begin{eqnarray}
0 &=& 2n L_0 |0 \ket + \frac{c}{12} n (n^2-1) |0 \ket \\
&=& [ L_n , L_{-n} ] | 0 \ket \\
&=& L_n L_{-n} |0 \ket - L_{-n} L_n |0 \ket \\
&=& L_n L_{-n} |0 \ket \, ,
\end{eqnarray}
and thus we have $L_{-n}|0\ket=0$ for all $n \in \mathbb{Z}$. This means that if we
expand $T(z)$ in powers of $z$, i.\,e.
\begin{equation}
T(z) = \sum_{n \in \mathbb{Z}} L_{n} z^{-n-2},
\end{equation}
we clearly see that $\bra 0 | T(z) = T(z) | 0 \ket = 0$ if we impose full 
conformal invariance which is possible if all states are orthogonal to $L_n|0\ket$
which is the case for the minimal model $c_{(3,2)}=0$. More precisely: the 
null vector is present in the irreducible vacuum representation but may
disappear in the full indecomposable representation based on $|\tilde{0}\ket$.
Note that if we include fields outside the Kac table without
assuming a Jordan cell structure for the identity level with $L_0 \lid = \id$,
non orthogonal states can be constructed since there are no constraints on their 
properties.

Nevertheless in our ansatz (the $c_{(9,6)}=0$ augmented minimal model), 
the state usually identified with the stress energy tensor seems to decouple
completely from the theory since it is even orthogonal to $l_{-2}|0\ket$ (there
may be additional $h=2$ fields present in the theory which are non orthogonal
but we do not know about any of them up to now).
However, this also forces any two point
function involving $T$ to vanish (as long as we do not modify the theory as
touched in section 3.3). Thus the first two-point 
function not to vanish is $\bra tt \ket$.

If we assume $L_{-2}|0\ket = T(0)|0\ket$  to be just an ordinary null state, 
$| \chi \ket$, and
not a fundamental property of the vacuum at $c_{(3,2)}=0$, we can obtain different results.

What we already know is that $L_{-2}|0\ket$ is a null state with respect to 
the action of all $L_n$. Thus we only have to check whether this holds for the
action of the $l_n$, too. Taking a look at 
\begin{equation}
\bra 0 |[l_2,L_{-2}]|0\ket = 4 \bra 0 |l_{0}|0\ket  + \mu \bra 0 ||0\ket = \mu \, ,
\end{equation}
we see that it is consistent to assume a non-orthogonal state to $L_{-2}|0\ket$ if
we exclude a Jordan cell for the identity. Even the Jordan cell relation between
the usual state associated with the stress energy tensor and its logarithmic partner
turns out to be as expected - 
$$
L_0 t(z)|0\ket\equiv L_0 l_{-2} |0\ket = 
2 l_{-2} |0\ket + L_{-2} |0\ket = 2 t(0)|0\ket + T(0)|0\ket \, .
$$
Once more we stress that $t(z)$ can only be non-orthogonal 
in a non Kac based approach to $c=0$ since otherwise
we know that we have a Jordan cell connection between the identity and other
states which would cause the two point function to vanish:
\begin{equation}
\bra 0 |[l_2,L_{-2}]|0\ket = 4 \bra 0 |l_{0}|0\ket  + \bra 0 |\mu \id + \frac{c}{2} \lid|0\ket = 0 \, .
\end{equation}

\subsection[Null vectors in a Kac table based $c=0$ theory]{Null vectors in a Kac table based \boldmath{$c=0$} theory}
Having agreed upon the proposal that the Kac-table of the augmented $c=0$ should be
taken, we know that under certain circumstances we can have null vectors in our
theory. The assumption that no other fields than those of the Kac-table and their
descendants may arise is crucial to this calculation since we do not have any
knowledge on the properties of non-Kac fields. Thus we point out that there are
problems with any arguments based on the
assumption of null vectors in a non strictly Kac based theory. 

Assuming $t(z)|0\ket$ to have a mode expansion like $T(z)|0\ket$, i.\,e.
\begin{equation}
t(z)|0\ket = \sum_{n \in \mathbb{Z}} l_n z^{-n-2}|0\ket \, ,
\end{equation}
and following the idea of Gurarie and Ludwig \cite{Gurarie:2004ce}, we will try to
construct universal null vectors that do not only vanish under the action of
all $L_n$ for $n >0$ but also after the application of $l_m$ for $m>0$. We have
to emphasize that this simple expansion of $t(z)$ only holds when acting on a 
highest weight state.
\subsubsection{The ordinary level two null vector}
Now let us have a look at the ordinary null vector on the second level
\begin{eqnarray} \label{eq:NV2}
| \chi_{h,c}^2 \ket &=& \left(L_{-2} - \frac{3}{2(2h+1)} L_{-1}^2 \right) |h\ket \, ,\\
h &=& \frac{1}{16}\left( 5 - c \pm \sqrt{(c-1)(c-25)} \right) \, .
\end{eqnarray}
What we already know is that $L_{\lbrace n \rbrace} | \chi_{h,c}^2 \ket = 0$
for all $|\lbrace n \rbrace| >0$ with 
$\lbrace n \rbrace = \lbrace n_1,n_2 \ldots, n_k \rbrace$ and 
$|\lbrace n \rbrace| = \sum_i n_i$. But what about the action of 
$l_{\lbrace n \rbrace }$ on $ | \chi_{h,c}^2 (0) \ket$? For $|\lbrace n \rbrace| >3$
this is obviously trivial since commuting the $l_{\lbrace n \rbrace}$ to the
right will leave us with some linear combination of $l_{\lbrace m \rbrace}$ and
$L_{\lbrace m' \rbrace}$ with $|\lbrace m \rbrace|,|\lbrace m' \rbrace| >0$ which
vanishes. Thus the interesting cases are the application of $l_2$ and $l_1^2$.

Therefore we have to use the algebra of the $L_n$ and $l_n$. The algebra between
the modes of $T(z)$,
\begin{equation} \label{eq:algebra0}
[L_n,L_m] = (n-m) L_{n+m}\, ,
\end{equation}
is just the same as in any ordinary CFT in spite of lacking the central extension 
due to the vanishing central charge. The mixed commutator is given by
\begin{equation} \label{eq:algebra01}
[L_n,l_m]|0\ket = (n-m) l_{n+m}|0\ket + (n+1) L_{n+m}|0\ket + \frac{\mu}{6}n (n^2-1) \delta_{n+m,0}|0\ket \, .
\end{equation}
Now we can test the known level two null vector from the ordinary
theory by applying the generators of $t(z)$:
\begin{equation} 
l_2 | \chi_{h,c}^2 \ket = \left[ \left(4 - \frac{18}{2(2h+1)}\right)l_0 \, .
+ h + \mu \right]| \chi_{h,c}^2 \ket.
\end{equation}
This result raises the question what the action of $l_0$ on some state of weight $h$ might be.
According to Gurarie and Ludwig \cite{Gurarie:2004ce}, it can be chosen to be 
equal to zero, $l_0 |h \ket = 0$, but this statement contradicts the level
one null vector assumption, i.\,e.\
\begin{equation} \label{eq:chi01}
\left( l_{-1} - \frac{1}{2} L_{-1} \right) | h \ket = 0
\end{equation}
for $h \neq 0$. To be sure, we prove this statement here. Thus let us 
have a look at this mixed level one null state from a general
point of view, i.\,e.\ $| \tilde{\chi}_{h,c}^1  \ket = \left( al_{-1} +b  L_{-1} \right) | h \ket $
for some suitable $|h\ket$. Claiming $L_n |\tilde{\chi}_{h,c}^1 \ket =0$ for all $n>0$
it follows that
\begin{equation} \label{eq:label}
\left( 2 a l_0 + 2 h(a+b) \right) | h \ket = 0 \, .
\end{equation}
Here we have to distinguish between different cases:
\begin{itemize}
\item[(a)] $h \neq 0$ :
\begin{equation} 
\begin{array}{rclcrcl}\label{eq:blitz}
l_0 |h\ket &=& 0 & \Rightarrow & a &=& -b \, ,\\
l_0 |h\ket &=& h |h\ket & \Leftrightarrow & a &=& -\frac{1}{2}b \, .
\end{array}
\end{equation}
Obviously the first result of (\ref{eq:blitz}) contradicts (\ref{eq:chi01}) but 
the second result is ok to agree upon.
\item[(b)] $h = 0$,  and thus
\begin{equation} 
L_n | \tilde{\chi}_{h,c}^1 \ket = a L_n l_{-1} |0 \ket = -a(n+1)l_{n-1} |0\ket
\end{equation}
which vanishes for $n > 0$.
\end{itemize}
Thus with reservation regarding the action of $l_n$ on
the mixed level one null state, $| \tilde{\chi}_{h,c}^1 \ket$ is a null vector
for all $h$, but with the special choice of \cite{Gurarie:2004ce} only for
$l_0 |h\ket = h |h\ket$. Conversely, if we say $l_0 |h \ket = 0$, we do not have
the special null vector (\ref{eq:chi01}) (independent on what the commutator
of the modes of $t(z)$ might be).

In spite of the fact that we do not know the exact form of the commutator 
$[l_n,l_m]$, we can conclude that $[l_1,l_{-1}] \propto a l_0 + b L_0 + \phi$
with $\phi$ being a linear combination of other primaries of weight zero. Thus for $h=0$ 
and the action of $l_0$ on $|h\ket$ either being $0$ or $h$, we can conclude that
$l_{-1}|0\ket = 0$ since
\begin{equation} 
l_1 | \tilde{\chi}_{h,c}^1 \ket = a l_1 l_{-1} | 0 \ket + b l_1 L_{-1} |0\ket 
= a [l_1,l_{-1}] |0 \ket + 2 b l_0 |0\ket = 0.
\end{equation}
Unfortunately, there is no criterion to draw the right conclusion which is the right
choice of the action of $l_0$ on a state. Thus we will just state
that for $l_0 |h\ket =h|h\ket$ we found $\mu=0$ for $h=0$ and $\mu = \frac{7}{8}$
for $h = \frac{5}{8}$ and for $l_0 |h\ket = 0$ we found $\mu=0$ for $h=0$ and 
$\mu = -\frac{5}{8}$ for $h = \frac{5}{8}$.

Although it is always consistent to assume the existence of the special 
null vector (\ref{eq:chi01}) if we choose $l_0 |h\ket =h|h\ket$, we 
do not know whether for $h \neq 0$ we have $l_1| \tilde{\chi}_{h,c}^1  \ket = 0$. 
But we have to be careful since this is only a circular reasoning. However, if we
assume the special choice of $| \tilde{\chi}_{h,c}^1 \ket$ as in (\ref{eq:chi01})
to be a null state, we encounter the fact that $l_1 | \chi_{h,c}^2 \ket$
is a true null state for the whole theory.
\subsubsection{The level three null vector}
We could try the same procedure on the level three null state
\begin{eqnarray}\label{eq:NV3}
| \chi_{h,c}^3 \ket &=& \left( L_{-1}^3 - 2\left(h + 1 \right) L_{-2} L_{-1} 
+ h\left(h + 1 \right) L_{-3}\right) \phi_{h} \, ,\\
h &=& \frac{1}{6}\left( 7 - c \pm \sqrt{(c-1)(c-25)} \right) \, .
\end{eqnarray}
Testing the level three null vector for consistency by applying $l_3$ to
$| \chi_{h,c}^3 \ket$, we find that
for $l_0 |h\ket = h |h\ket $ and $h = \frac{1}{3}$ we get 
$\mu= - \frac{h(h-2)}{2(h-1)} = - \frac{5}{12}$; for $h=2$ we get $\mu = 0$.
For $l_0 |h\ket = 0 $ and $h = \frac{1}{3}$ we get $\mu= 6-2h = \frac{17}{3}$
and for $h=2$ we get $\mu = 2$. We have to bear in mind that these are only
necessary conditions, we did not check whether the action of $l_2 l_1$ and $l_1^3$
supports this result or gives a contradiction, meaning that there is no level
three null state in the theory any more.
\subsubsection[Comments on percolation as an augmented $c=0$ model]{Comments on percolation as an augmented \boldmath{$c=0$} model}
Independently we can conclude that we do not only have different theories for 
different values of $\mu$ but also that any given $c=0$ theory splits up in 
certain subsets of primary operators which ''cannot give rise to [...] differential
equations simultaneously in the same theory'' \cite{Gurarie:2004ce}.

This remark alone shows that, in the case of percolation, it is not
sensible to try to derive a second order differential equation for the horizontal
crossing probability $\Pi_h$ and a third order differential equation for the
horizontal-vertical crossing probability $\Pi_{hv}$ in the same Kac table based theory. Moreover,
we do not yet know if such a level three null state even exists. Thus this may be
another hint that $c=0$ for percolation may not be the correct choice
\cite{Flohr:2005ai}.

\subsection[The field content of a $c_{(9,6)}=0$ augmented minimal model]{The field content of a \boldmath{$c_{(9,6)}=0$} augmented minimal model}
\enlargethispage{0.5cm}
After having talked so much about the augmented $c_{(9,6)}=0$ model,
we should give at least a brief overview on its features since there
has been not much literature published about generalized augmented $c_{(p,q)}$
models with $q > 1$ so far. Following the ideas of \cite{Flohr:1996vc},
we know that the smallest closing set of modular functions 
larger than the $\frac{1}{2}(p-1)(q-1)$ characters for the minimal
$c_{(p,q)}$ model contains $\frac{1}{2}(3p-1)(3q-1)$ 
individual functions which stay in some suitable linear combination
in direct correspondence
to the number of highest weight representations or fields in the 
augmented Kac table.
The modular functions can be found by solving the modular differential equation
as introduced in \cite{Mathur:1988gt,Mathur:1988na}. The generalization of this 
method towards LCFT can be found in \cite{Flohr:2005cm}.
In our example, the $c_{(9,6)}$ model, twenty torus amplitudes can be 
matched with the
twenty representations of the modular group being present in the Kac
table of $c_{(9,6)}=0$ \cite{Holger}.
Closed sets of such functions can only be obtained considering an odd multiple
of $(p,q)$ thus usually one tries to get along with the smallest set, i.\,e.\ $(3p,3q)$. 

Thus in contrary to the minimal model $c_{(p,q)}$ we technically have to deal
with an extended Kac table of $c_{(3p,3q)}$: 
\begin{equation} \label{eq:Kac}
c_{(9,6)}\quad : \quad
\begin{array}{|c|c|c|c|c|c|c|c|} 
\hline
\rule[-.6em]{0cm}{1.6em}0& 0 &\frac{1}{3}&1 &2 &\frac{10}{3}& 5 & 7\\
\hline
\rule[-.6em]{0cm}{1.6em} \frac{5}{8} & \frac{1}{8} & -\frac{1}{24} & \frac{1}{8} & \frac{5}{8} & \frac{35}{24} & \frac{21}{8} & \frac{33}{8}\\
\hline
\rule[-.6em]{0cm}{1.6em} 2 & 1 & \frac{1}{3} & 0 & 0 & \frac{1}{3} & 1 & 2\\
\hline
\rule[-.6em]{0cm}{1.6em} \frac{33}{8} & \frac{21}{8} & \frac{35}{24} & \frac{5}{8} & \frac{1}{8} & -\frac{1}{24} & \frac{1}{8} & \frac{5}{8} \\
\hline
\rule[-.6em]{0cm}{1.6em} 7 & 5 & \frac{10}{3} & 2 & 1 & \frac{1}{3} & 0 & 0 \\
\hline
\end{array}\quad .
\end{equation}
As it is always the case for $c_{(3p,3q)}$ augmented models, we have
$3\times 2$ fields in the Kac table which are of weight $h=0$ and lie
within the upper left and lower right corners of the replicated minimal
Kac tables on the diagonal. It is conjectured \cite{Holger} that all fields 
inside the boundary of the replicated minimal Kac table belong to rank $3$
Jordan cells whose detailed structure is not yet known.

Fields on the boundary of the replicated minimal Kac table show up with
a multiplicity of $2 \times 2$ and belong to rank $2$ Jordan cells. The 
corresponding representation of weight $h_{(r,s)}+rs$
is present $1 \times 2 $ times as expected, too. Additionally, the fields on the edges
of the boundaries show up only $1 \times 2$ times as well, with their
corresponding representations of weight $h_{(p,q)} + pq /4$ showing up
at the anti-diagonal edges.

Thus in the special case of $c=0$ we have
two highest weights which do not form Jordan cells, i.\,e.\
$-\frac{1}{24},\frac{35}{24}$ while the other operators of the boundary of the
conformal grid are arranged in triplets of which two states of the
same weight  form an indecomposable representation 
and one belongs to an irreducible representation which 
is differing by an integer in its weight (more precisely $rs$), i.\,e.\
\small
\begin{eqnarray}
\left( \frac{5}{8} ,\frac{5}{8} ,\frac{21}{8} \right)
&=&\left( \frac{5}{8} ,\frac{5}{8} ,2 + \frac{5}{8} \right) \, ,\\
\left( \frac{1}{3} ,\frac{1}{3} ,\frac{10}{3} \right)
&=&\left( \frac{1}{3} ,\frac{1}{3} ,3 + \frac{1}{3} \right)\, ,\\
\left( \frac{1}{8} ,\frac{1}{8} ,\frac{33}{8} \right)
&=&\left( \frac{1}{8} ,\frac{1}{8} ,4 + \frac{1}{8} \right)\, .
\end{eqnarray}
\normalsize
Due to these indecomposable representations, logarithms arise in the OPEs and 
especially in the fusion product of the pre-logarithmic field 
$\phi_{-\frac{1}{24}}$ with itself. 

The sector containing the $h=0$ fields has a more complicated structure. We have
three multiple weights $(0,0,0), (1,1)$ and $(2,2)$ but we do not yet 
know how they are arranged among the other two fields of weights $5$ and $7$, 
respectively. As stated above, it is conjectured \cite{Holger} that
they may form a rank three Jordan cell structure whose details are
currently being worked out. Additionally, we can not exclude exotic behavior
such as Jordan cells with respect to other generators than $L_0$, e.\,g.\
$\mathcal{W}$-algebra zero modes. Even worse, there might exist indecomposable
structures with respect to $L_n$, $n\neq 0$, as in \cite{Krohn:2002gh}.

As far as we know there has not been any research concerning this issue 
before. It seems reasonable to assume a structure related to that of the $c_{(p,1)}$
models which has already been discussed in detail \cite{Gaberdiel:1996np}, 
\cite{Gaberdiel:1998ps}, \cite{Kausch:2000fu}, but obviously at least for the 
integer weights it can not be the whole story.

If we accept that the Kac-table of $c=0$ has to be extended beyond its
minimal truncation, we immediately encounter a problem. The field corresponding
to the entry $(2,3)$ in the Kac-table has a negative conformal
weight $h_{2,3} = -1/24$. Hence, the theory cannot be unitary. Furthermore,
the effective central charge $c_{\mathrm{eff}}= c - 24h_{\mathrm{min}}$ with
$h_{\mathrm{min}}$ the minimal eigen value of $L_0$ is then given by
$c_{\mathrm{eff}}=(c=0)-24(h=-1/24) = 1$. It follows that such a theory
cannot be rational with respect to the Virasoro algebra alone, but only
quasi-rational. However, there presumably exists an extended chiral symmetry 
algebra, ${\cal W}(2,15,15,15)$ under which the theory is rational 
\cite{Holger}. Fortunately, most of the structures which will interest us in
this paper can be studied from the the perspective of the Virasoro algebra.

As a concluding remark, let us note that
there seems to be a connection to $c_{(6,1)}=-24$ which is the only rational
(L)CFT with equal central charge modulo $24$ and thus exhibiting the same modular
properties. This theory also has effective central charge one.
Unfortunately, the analogies only hold for the boundary of the Kac
table and therefore we can only deduce the properties for the 
representations from the boundary of the Kac-table
of the $c=0$ model and not for the integer weight states.

\section{The forgotten loophole}
\subsection[Tensorized (L)CFTs with $c=0$]{Tensorized (L)CFTs with \boldmath{$c=0$}}
Obviously there is a fourth way out of the dilemma. Taking two non-interacting
CFTs with central charges $c_1$ and $c_2=-c_1$, respectively, and tensorizing them,
we get a CFT with vanishing central charge again but the OPE (\ref{eq:OPE}) looks like
\small
\begin{equation}
\phi_h(z) \phi_h^\dagger(0) \sim \frac{C^{\id}_{\Phi\Phi}}{z^{2h}}\left( 1 + \frac{2h}{c_1} z^2 
(T_{c_1}(0)- T_{-c_2}(0)) + \ldots \right) + \ldots \quad ,
\end{equation}
\normalsize
which is perfectly well defined for $c = 0$ if $c_1 \neq 0$.

But the result comes with a price, too: we have to introduce a new field $t(z) := 
T_{c_1}(z)-T_{-c_2}(z)$ which can be shown to satisfy the
following OPEs with the stress energy tensor \cite{Gurarie:2004ce} 
\begin{eqnarray}
T(z)T(0) &\sim& \frac{2 T(0)}{z^2} + \frac{ T'(0)}{z} + \ldots \quad ,\\ \label{eq:logOPE}
T(z)t(0) &\sim& \frac{c_1}{z^4} + \frac{2 t(0)}{z^2} + \frac{ t'(0)}{z} + \ldots \quad ,\\
t(z)t(0) &\sim& \frac{2 T(0)}{z^2} + \frac{ T'(0)}{z} + \ldots \quad .
\end{eqnarray}

The OPE of the tensorized $c=c_1+c_2=0$ LCFT model consists 
of an ordinary CFT part from the $c_2$-sector and a LCFT part from the $c_1$ sector. 
Thus we would get a $c=0$ theory with logarithmic operators without vanishing two-point
function.

Operators in the full tensorized theory therefore are just direct products
$\phi_h^{(0)} = \phi_{h_1}^{(1)} \otimes \phi_{h_2}^{(2)}$ whose weights are given by the sum of
both parts $h = h_1 + h_2$. Thus the OPE of a primary field is given by (see \cite{Kogan:2001ku})
\begin{eqnarray*}
\phi_h^{(0)} (z) \phi_h^{(0)} (0) &=& \phi_{h_1}^{(1)} (z) \phi_{h_1}^{(1)} (0) \otimes 
\phi_{h_2}^{(2)} (z) \phi_{h_2}^{(2)} (0) \\
&\sim& \frac{1}{z^{2h_1}} \left(\id^{(1)} + z^2 \frac{2 h_1}{c_1} T^{(1)}(0) + \ldots \right)\\
&\quad& \quad \quad \quad \quad \times \quad
\frac{1}{z^{2h_2}} \left(\id^{(2)} + z^2 \frac{2 h_2}{c_2} T^{(2)}(0) + \ldots \right) + \ldots \\
&\sim& \frac{1}{z^{2h}} \left( 1 + z^2 \left(\frac{2 h_1}{c_1} T^{(1)}(0) + 
\frac{2 h_2}{c_2} T^{(2)}(0)\right) \right)\, ,
\end{eqnarray*}
which is well defined since the $c_i \neq 0$ and the theories by themselves are regular.
\subsection{The general case}
In some cases we may not be able to choose a (bosonic) free field construction for the 
stress-energy-tensor. Thus we have to take a look at the general OPEs for a tensorized theory of an 
LCFT with central charge $c_1$ and an ordinary CFT with $c_2 = - c_1$. We start with the known OPEs
\begin{equation}\label{eq:OPEansatz}
\begin{array}{rcl}
T^{(i)}(z) T^{(i)}(w) &=& \frac{\frac{c_i}{2}}{(z-w)^4} + \frac{2 T^{(i)} (w)}{(z-w)^2} + \frac{\partial_w T^{(i)} (w)}{(z-w)}\, ,\\
\lid^{(1)} (z) \lid^{(1)}(w) &=& \log^2 (z-w)\id^{(1)} + 2 \lzw \lid^{(1)} (w)\, ,\\
T^{(1)} (z) \lid^{(1)}(w) &=& \frac{\id^{(1)}}{(z-w)^2} + \frac{\partial_w 
\lid^{(1)} (w)}{(z-w)}\, ,
\end{array}
\end{equation}
and we define
\begin{eqnarray*}
t^{(1)} (w)&:=& \, \nop{ T^{(1)}\lid^{(1)} } (w)\, ,\\
t^{(0)} (w)&:=& t^{(1)} (w)\otimes \id^{(2)} + (\alpha \id^{(1)} + \beta 
\lid^{(1)}(w)) \otimes T^{(2)}(w)\, .
\end{eqnarray*}
To obtain the two point functions, we make the ansatz:
\begin{equation}\label{eq:ansatz}
\begin{array}{rcl}
T^{(0)}(z) &=& T^{(1)}(z) \otimes \id^{(2)}(z) + \id^{(1)}(z) \otimes T^{(2)}(z) \, \\
t^{(0)}(z) &=& t^{(1)}(z) \otimes \id^{(2)}(z) + (\alpha \id^{(1)}(z) + \beta \lid^{(1)}(z)) \otimes T^{(2)}(z)\, .
\end{array}
\end{equation}
This leaves us with the following results for 
$T^{(0)} (z) T^{(0)}(w)$ and $T^{(0)} (z) t^{(0)}(w)$:
\begin{eqnarray*}
T^{(0)} (z) T^{(0)} (w) 
&=& T^{(1)} (z) T^{(1)} (w)  + T^{(2)} (z) T^{(2)} (w) \\
&\sim& \frac{\frac{c_1}{2}+\frac{c_2}{2}}{(z-w)^4} + 
\frac{2 (T^{(1)} + T^{(2)}) (w)}{(z-w)^2} + \frac{\partial_w (T^{(1)} + T^{(2)}) (w)}{(z-w)}\\
&=& \frac{2 T^{(0)}(w)}{(z-w)^2} + \frac{\partial_w T^{(0)} (w)}{(z-w)}\, ,
\end{eqnarray*}
whereas the OPE with its logarithmic partner
\begin{eqnarray*}
T^{(0)} (z) t^{(0)} (w) 
&=& T^{(1)} (z) t^{(1)} (w)\otimes\id^{(2)} + (\alpha\id^{(1)} + \beta \lid^{(1)})\otimes T^{(2)} (z) T^{(2)} (w) \\
&\sim& \frac{\frac{c_1}{2}\left((1-\beta)\lid^{(1)}-\alpha\id^{(1)} \right)\otimes\id^{(2)}}{(z-w)^4} + \frac{2 t^{(0)} (w) + T^{(1)} (w)\otimes\id^{(2)}}{(z-w)^2} + \frac{\partial_w t^{(0)} (w)}{(z-w)}
\end{eqnarray*}
yields a non-vanishing vev with a modified $b$-term:
\begin{eqnarray*}
\bra T^{(0)} (z) t^{(0)} (w) \ket &=& \frac{\frac{c_1}{2}(1-\beta)}{(z-w)^4}\, .
\end{eqnarray*}
For the OPE of the logarithmic partner fields, we get
\small
\begin{eqnarray*}
&&t^{(0)} (z) t^{(0)} (w)\\ 
&\sim&  \frac{1}{(z-w)^4} \left(\left(1 + \alpha^2 \frac{c_2}{2}\right) + \left(\frac{c_1}{2} + \beta^2 \frac{c_2}{2}\right)\log^2 (z-w) \right.\\ 
&\quad& \left.- 2\left(\frac{c_1}{2} + \beta^2 \frac{c_2}{2}\right) \lzw \lid (w)+ \alpha \beta \lid c_2 + \left(\frac{c_1}{2} + \beta^2 \frac{c_2}{2}\right)\nop{\lid(z)\lid_1(w)}  \right)\\
&\quad& +\frac{1}{(z-w)^2} \left( 2\left( T^{(0)}(w) - (1-\beta^2) T^{(2)}(w) \right)\log^2 (z-w) \right.\\
&\quad&- 4\left( t^{(0)} (w)- \alpha \beta \lid T^{(2)}(w) \right)\lzw + 2 t^{(0)}(w) + 2 \alpha \beta \lid T^{(2)}(w) \\
&\quad& \left.+  \left( T^{(0)}(w) - (1-\beta^2) T^{(2)}(w) \right)\nop{\lid(w) \lid(w)} \right)\\
&\quad& + \frac{1}{(z-w)} \left(\left(\del T^{(0)}(w) - (1-\beta^2) \del T^{(2)}\right)\log^2 (z-w) \right.\\
&\quad& - 2\left( \del t^{(0)}(w) - \alpha \beta \lid \del T^{(2)}(w) \right)\lzw \\
&\quad& \left.+ \del t^{(0)} (w)+  \alpha \beta \lid \del T^{(2)}(w)  + \del \left(\left(T^{(0)}(w) - (1-\beta^2) T^{(2)}(w)\right)\nop{\lid(w) \lid(w)}\right) \right)\\
&\quad& + \left(\log^2 (z-w) - 2 \lzw \lid (w)\right)\\
&\quad& \quad \quad \quad \quad \times \left(\nop{T^{(0)}(w)T^{(0)}(w)} - (1-\beta^2)\nop{T^{(2)}(w)T^{(2)}(w)}\right)\, ,
\end{eqnarray*}
\normalsize
where we suppressed the labels for the tensor factors as they are clear from
the context.
Obviously the only possibility to get nothing but ''zero charge'' quantities on 
the rhs is to put $\alpha=0$ and $\beta=1$ which means, that we are left with vanishing 
vevs for $\bra TT \ket$ and $\bra Tt \ket$. In that case the equations would be of the 
same form as for the ordinary $c=0$ LCFT and our construction would be useless. To be
exhaustive, we will give the vev of this calculation, too,
\begin{eqnarray*}
\bra t^{(0)}(z) t^{(0)}(w) \ket  &=&c_2\frac{ \alpha \beta + \left(1 - \beta^2 \right) \lzw }{(z-w)^4}\, .
\end{eqnarray*}

%
\subsection{An example of a tensor model}
One of many possible applications
is a tensor product of a $c=-2$ theory and four Ising models. This ansatz has many 
advantages, e.\,g.\ a logarithmic pair in the identity sector of the part with $c=-2$
and the closure under fusion of a small subset of the fields. 

As stated in \cite{Kogan:2001nj} and \cite{Kogan:2001ku}, this corresponds to an $SU(2)_0$
or $OSp(2|2)_{-2}$ model, where the logarithmic structure appears in the $c=-2$ part. 

\paragraph{The Ising Model} In the Ising Model, we have the following fields
\begin{eqnarray}
\id &= \phi_{(1,1)}, \phi_{(2,3)} \quad &h = 0 \, ,\\
\s &= \phi_{(2,2)}, \phi_{(1,2)} \quad &h = \frac{1}{16}\, ,\\
\e &= \phi_{(2,1)}, \phi_{(1,3)} \quad &h = \frac{1}{2}\, ,
\end{eqnarray}
with the following fusion rules:
\begin{eqnarray}
\s \times \s &=& \id + \e \, ,\\
\s \times \e &=& \s\, ,\\
\e \times \e &=& \id \, .
\end{eqnarray}
For $c=c_{2,1}=-2$, we have an indecomposable representation of the 
$h=0$ sector ($\mathcal{R}_{\id}$) consisting of two fields with $h=0$ whose details are not important for our further discussion and two others, i.\,e.\ $\mu$ with $h=-\frac{1}{8}$ and $\nu$ with $h=\frac{3}{8}$. These fields obey 
\begin{eqnarray}
\mu \times \mu = \mu \times \nu = \nu \times \nu &=& \mathcal{R}_{\id}\, ,\\
\mu \times \mathcal{R}_{\id} = \nu \times \mathcal{R}_{\id} &=& \mu + \nu \, ,\\
\mathcal{R}_{\id} \times \mathcal{R}_{\id} &=& 2 \mathcal{R}_{\id} \, .
\end{eqnarray}
It is easy to check, that the symmetrized fields of the four Ising models 
$\id, E_1, E_2, E_3, E_4$ and $S$ (where $E_i$ denotes the totally symmetric
tensor product of $i$ fields $\e$ and $4-i$ fields $\id$ and 
$S=\otimes^4 \s \equiv (\s,\s,\s,\s)$) close under fusion. From 
these fields tensorized with those of $c=-2$ we can choose a consistent subset 
$(\mathcal{R}_{\id},\id)$, $(\mathcal{R}_{\id}, E_i)$, $(\mu, S)$ and $(\nu, S)$. Obviously, 
$(\nu,S)$ has conformal weight $h= \frac{5}{8}$ and $(\mu,S)$ has conformal 
weight $h= \frac{1}{8}$ which are fields assumed to appear in percolation.
However, if percolation can be described by a $c=0$ model such as $(c=2) \otimes
(c=-2)$, the question remains how Watts' differential equation  \cite{Watts:1996yh} 
can be derived through a level three null vector condition acting on a four point 
function of boundary changing operators in this theory \cite{Flohr:2005ai}.

\paragraph{The operator product expansion} 
The OPE of the tensorized $c=0$ model, 
consists of an ordinary CFT part from the $c=2$ sector and and LCFT part from the 
$c=-2$ sector. To obtain the two point functions, we make the same ansatz as before
(\ref{eq:ansatz}) 
\begin{eqnarray}
T(z) &=& \nop{\del \theta^+(z) \del \theta^-(z) }\, ,\\
\lid (z) &=& \nop{\theta^-(z) \theta^+(z)}\, , \\
t(z) &=& \nop{T(z) \lid (z)}\, .
\end{eqnarray}

The results for $\bra Tt \ket$ and $\bra Tt \ket$ are exactly the
same as for the general case. Since the OPE of $t(z)t(w)$ is relatively short, we 
will state all terms:
\small
\begin{eqnarray*}
&&t^{(0)} (z) t^{(0)} (w)\\ 
&=& t^{(1)}(z) t^{(1)}(w) + \left(\alpha + 2 \alpha \beta \lid + \beta^2 \lid(z) \lid (w) \right) T^{(2)}(z) T^{(2)}(w)\\
&\sim& \frac{1}{(z-w)^4}\left( \log^2 (z-w) + 2\lzw \lid (w) + 1 \right)\\ 
&\quad& +\sum_{i=0}^{3}\frac{\lzw \del^i\lid(w) }{i!(z-w)^{4-i}}+\frac{1}{2(z-w)}\del^3 \lid(w)\\
&\quad& +\frac{1}{(z-w)^2}\left( \left[\lzw - 2 \log^2 (z-w) \right] T^{(1)}(w) + \left[2 - 4 \lzw \right] t^{(1)}(w)\right)\\
&\quad& +\frac{1}{2(z-w)}\left( \left[\lzw - 2 \log^2 (z-w) \right] \del T^{(1)}(w) + \left[2 - 4 \lzw \right] \del t^{(1)}(w)\right) \\
&\quad& +\frac{\frac{c_{2}}{2}(\alpha^2 + 2 \alpha \beta \lid - \beta^2 (2\lzw \lid + \log^2(z-w))}{(z-w)^4} \\
&\quad& + \frac{2(\alpha^2 + 2 \alpha \beta \lid - \beta^2 (2\lzw \lid + \log^2(z-w)) T^{(2)} (w)}{(z-w)^2} \\
&\quad& + \frac{\partial_w [(\alpha^2 + 2 \alpha \beta \lid - \beta^2 (2\lzw \lid + \log^2(z-w))T^{(2)} (w)]}{(z-w)}\, .
\end{eqnarray*}
\normalsize
Note that since the $\theta$ anti-commute, $\nop{ \lid(w) \lid(w)}$ vanishes.

Obviously here, too, it is not possible to reduce the rhs of the equation to 
terms only consisting of the 'neutral' operators, since it would be necessary to 
set $\alpha=0$ and $\beta = 1$ which means that the OPEs of $T^0 (z) t^0(w)$ 
would vanish. Even the vev of the two-point function of the logarithmic partner 
vanishes in this case:
\begin{eqnarray}
\bra t^{(0)} (z) t^{(0)} (w) \ket &=& c_2\frac{\alpha \beta  + 2 (1- \beta^2 ) \lzw}{(z-w)^4}\, .
\end{eqnarray}
\section{Concluding remarks}
\enlargethispage{0.5cm}
In our investigation of the structure of (L)CFTs with
vanishing central charge we chose a new approach based on the
augmented minimal model $c_{(9,6)}=0$, including a Jordan cell
structure on the identity level with respect to $L_0$. From 
this assumption follows immediately the Jordan cell connection
of the level two descendants of $\id$ and its logarithmic 
partner $\lid$, $L_0 t(z) = 2 t(z) + \lambda T(z)$.
A special feature of this setup is the vanishing of any
two point function involving $T(z)$. Depending on the different resolutions
of this puzzle, one is forced to take certain consequences into account.
If we stick to taking $T(z)|0\ket\equiv 0$ in the irreducible vacuum
representation of the Virasoro algebra, we might
reconsider the field state isomorphism for $c=0$. Also, the assumption of
a logarithmic partner of the identity naturally leads to a vanishing of
all correlation functions which only involve proper primary fields. 
The stress energy tensor is a proper primary field in the case of vanishing
central charge. 

A possible way out of this dilemma is to interpret the vanishing of
correlators as being due to the presence of certain zero modes. Such
behavior is well known in fermionic theories such as ghost systems and
particularly $c=-2$. Only when the zero modes are canceled due to certain
field insertions do we get non-vanishing results. It would be most 
tempting to try to construct a free field realization of a $c=0$ theory
as a Kac-table based theory with anti-commuting fields. 

We presented several arguments why the Kac table based ansatz
is more promising than the replica approach, especially with respect
to its interpretation in physics and determination of the field
content. Furthermore we gave a fourth loophole to the 
$c \rightarrow 0$ catastrophe within an (L)CFT setup and
gave examples for both approaches to $c=0$. 

Since none of the approaches to $c=0$ currently seems to be
able to fulfill all wanted features by now at a time we suggest 
further investigation. This includes
a fermionic realization of the augmented minima model and, 
above all, general research on the representation theory of the 
augmented minimal $c_{(9,6)}=0$ model with extended Kac table.
Thus with our ansatz, we therefore discovered an interesting application for
augmented minimal models, or, more precisely an extension of
the (L)CFT formalism of $c_{(p,1)}$ models to $c_{(p,q)}$ with arbitrary 
$q \in \mathbb{Z}_+$. Hence, the investigation of the representations
inside the boundary of the original replicated Kac table will be an important field
of research in the future \cite{Holger}.

We believe that the results of this paper have been a small but important step
towards the implementation of percolation models within an
LCFT approach. We have shown that we should reconsider widely accepted
assumptions such as the postulation of $c=0$ for percolation and Jordan cell
structures on higher level without the same structure for
the identity in $c=0$ theories. More precisely, we proved that there can not be
a level three and level two null vector condition in a $c=0$
augmented minimal model simultaneously for a standard assumption of
the action of $l_0$. This would exclude
either Watts' or Cardy's differential equation for the
two crossing probabilities in percolation.
\subsection*{Acknowledgment}
\enlargethispage{0.5cm}
We would like to thank Holger Eberle and 
Alex Nichols for helpful discussions and comments. 
The research of M.F.\ is partially supported by the European Union Network 
HPRN-CT-2002-00325 (EUCLID).
\appendix 
\section{Appendix}
\subsection[The algebra between the modes of $T(z)$ and $t(z)$]{The algebra between the modes of \boldmath{$T(z)$} and \boldmath{$t(z)$}}
What remains to be shown is that the algebra (\ref{eq:algebra01}) is correct
although it differs from the one given in \cite{Gurarie:2004ce}. We use the
general ansatz for a logarithmic field, 
\begin{eqnarray}\label{eq:modelog}
\tilde{\phi}_h (z) &=& \sum_{m \in \mathbb{Z},q \in \mathbb{N}_0} l_{m,q} \log^q(z) z^{-m-h}\, ,
\end{eqnarray}
to calculate the commutator by comparison of the powers of $\log (w)$ and $w$ on 
both sides of the equation. Note that this method is only applicable to the
mixed algebra and the ordinary between the $L_n$ alone since otherwise the
residue theorem would have to be applied to a non analytic function, i.\,e.\ $\log (w)$.
\small
\begin{eqnarray*}
&&[L_n,t(w)]\\
&=&\sum_{{m \in \mathbb{Z} \atop q \in \mathbb{N}_0}} [L_n,l_{m,q}] \log^q (w) w^{-m-2}\\
&=&\oint_0  \mathrm{d}z\, z^{n+1} T(z) t(w)\\
&=&\oint_0  \mathrm{d}z\, z^{n+1} \left(\frac{c/2\lid + \mu \id}{(z-w)^4} 
           + \frac{2t(w) + \lambda T(w)}{(z-w)^2}
           + \frac{\partial t(w)}{(z-w)^{1}}\right)\\
&=& w^{n-2} \frac{n(n^2-1)}{6}\left(c/2 \lid + \mu \id \right)	
    + (n+1) w^n\left( 2t(w) + T(w) \right) + w^{n+1} \del_w t(w) \\	 
&=& \sum_{{m \in \mathbb{Z} \atop q \in \mathbb{N}_0}} \log^q (w) w^{-m-2}
    \left[ \left( \frac{n(n^2-1)}{6}\left(c/2 \lid + \mu \id \right)\delta_{n+m,0} 
	      +(n+1)L_{n+m} \right) \delta_{q,0} \right.\\
&\quad& \quad \quad \quad \left. (n-m)l_{n+m,q} + (q+1)l_{n+m,q+1} { \, \atop \, }\right]\, .	 
\end{eqnarray*}
\normalsize
from which we may extract the commutator
\begin{eqnarray}\label{eq:Lnlmcomp}
[L_n,l_{m,}] &=& \left( \frac{n(n^2-1)}{6}\left(c/2 \lid + \mu \id \right)\delta_{n+m,0} 
	      +(n+1)L_{n+m} \right) \delta_{q,0} \nonumber\\
&\quad& +(n-m)l_{n+m,q} + (q+1)l_{n+m,q+1}\, .
\end{eqnarray}
A rather practical than elegant way out of the problem of complicated
commutators as (\ref{eq:Lnlmcomp}) is the application of the whole thing
to the vacuum (or any other highest weight state). 
Imposing regularity at $w \rightarrow 0$
we can conclude, that all modes with $q \neq 0$ have to vanish in that case.
Thus we are left with an analytic expression for $t(w)$, i.\,e.\
\begin{eqnarray}
t(w) |0\ket = \sum_{m \in \mathbb{Z} } l_m w^{-m-2}|0\ket  \, ,
\end{eqnarray}
and we could even calculate the OPE as in the usual way for non logarithmic
fields, 
\small
\begin{eqnarray*}
&&[L_n,l_m]|0\ket   \\
&=& \frac{1}{(2 \pi\mathrm{i})^2} \oint_0  \mathrm{d}w\, w^{m+1} \oint_0  \mathrm{d}z\, z^{n+1} 
\left(\frac{c/2\lid + \mu \id}{(z-w)^4} 
           + \frac{2t(w) + \lambda T(w)}{(z-w)^2}
           + \frac{\partial t(w)}{(z-w)^{1}}\right)|0\ket \\
&=& \frac{1}{2 \pi \mathrm{i}} \oint_0  \mathrm{d}w\, w^{m+n-1}
	\frac{n(n^2-1)}{6}(\frac{c}{2}\lid + \mu \id)|0\ket \\
&\quad& +\frac{1}{2 \pi \mathrm{i}} \oint_0  \mathrm{d}w\, w^{m+n+1}
		(n+1)(2t(w)+\lambda T(w))|0\ket 
		+ \frac{1}{2 \pi \mathrm{i}} \oint_0  \mathrm{d}w\, w^{m+n+2}\partial t(w)|0\ket \\
&=& \left( \frac{n(n^2-1)}{6}(\frac{c}{2}\lid + \mu \id) \delta_{n,-m} +
    (n+1)(2l_{n+m} + \lambda L_{n+m})  - (n+m+2) l_{n+m}\right)|0\ket \, ,
\end{eqnarray*}
\normalsize
and therefore
\begin{equation}\label{eq:Lnlm}
[L_n,l_m]|0\ket  = (n-m)l_{n+m}|0\ket  + (n+1) \lambda L_{n+m}|0\ket\, ,
+ \frac{n(n^2-1)}{6}\mu \delta_{n+m,0}|0\ket
\end{equation} 
with $T(z)t(w)$ as given in (\ref{eq:Tt}). 
%
\subsection[Mode expansion of $t(z)$ in $c=-2$]{Mode expansion of \boldmath{$t(z)$} in \boldmath{$c=-2$}}
As a concrete example for a non trivial mode expansion containing
logarithms, we chose the $c=-2$ CFT which is known to have a special 
realization containing 
\begin{equation}\label{eq:mode-2}
\theta^\pm = \theta_0^\pm \log(z) + \xi^\pm + \sum_{n \neq 0} \theta_n^\pm z^{-n}\, ,
\end{equation}
with the modes of $\theta^\pm$ obeying the canonical anti-commutation relations:
\begin{eqnarray}\label{eq:anticom-2}
\lbrace \theta_n^\pm , \theta_m^\mp  \rbrace &=& \frac{1}{n} \delta_{m+n,0}\, , \\ 
\lbrace \xi^\pm , \theta_0^\mp  \rbrace &=& \pm 1 \, .
\end{eqnarray}
The logarithmic partner of the identity is given by
\begin{eqnarray}\label{eq:modelid}
\lid(z) &:=& \nop{\theta^- \theta^+}(z) = \sum_{n \in \mathbb{Z}} \left( \imath_n + \log(z)\tilde{\imath}_n + \log^2(z)\hat{\imath}_n \right) z^{-n}\,.
\end{eqnarray}
Inserting (\ref{eq:mode-2}), we observe
\small
\begin{eqnarray*}
\lid(z) &:=& \nop{\theta^- \theta^+}(z)\\
&=& \nop{\left( \theta_0^- \log(z) + \xi^- + \sum_{n \neq 0} \theta_n^- z^{-n}\right)
    \left( \theta_0^+ \log(z) + \xi^+ + \sum_{m \neq 0} \theta_m^+ z^{-m}\right)}\\
&=& \log^2(z)\nop{\theta^-_0 \theta^+_0}  \\
&\quad& + 
    \log(z)\left[\left(\nop{\theta^-_0 \xi^+} + \nop{\xi^- \theta^+_0} \right) + 
    \left(\sum_{n \neq 0} \left(\nop{\theta^-_n \theta^+_0} + \nop{\theta^-_0 \theta^+_n}\right) z^{-n} \right)\right]\\
&\quad& + 
    \nop{\xi^- \xi^+}  + \sum_{n \neq 0} \left[\nop{\theta^-_n \theta^+_{-n}} + 
	\left(\nop{\xi^- \theta^+_n} + \nop{\theta^-_n \xi^+} \right)z^{-n} \right]\\
&\quad& \quad\quad\quad\quad+ 	
    \sum_{m,n \neq 0 \atop n\neq m}\nop{\theta^-_n \theta^+_{-m}}z^{m-n}\, .
\end{eqnarray*}
\normalsize
Now we can identify the terms. For $n\neq 0$ the modes of $\lid(z)$ are
\begin{eqnarray*}
\imath_0 &=& \nop{\xi^- \xi^+}  + \sum_{n \neq 0}\nop{\theta^-_n \theta^+_{-n}}\, ,\\
\imath_n &=& \sum_{n \neq 0}\left(\nop{\xi^- \theta^+_n} + \nop{\theta^-_n \xi^+} \right)z^{-n} + 
    \sum_{n,m \neq 0, n\neq m}\nop{\theta^-_n \theta^+_{-m}}z^{m-n}\, ,\\
\tilde{\imath}_0 &=& \nop{\theta^-_0 \xi^+} + \nop{\xi^- \theta^+_0}\, ,\\
\tilde{\imath}_n &=& \sum_{n \neq 0} \left(\nop{\theta^-_n \theta^+_0} + \nop{\theta^-_0 \theta^+_n}\right) \, ,\\
\hat{\imath}_0 &=& \nop{\theta^-_0 \theta^+_0}\, ,\\
\hat{\imath}_n &=& 0 \, .
\end{eqnarray*}
Since $t(z)= \, \nop{ T(z) \lid(z)}$ with $T(z) = \, \nop{\del \theta^+(z) \del \theta^-(z) }$,
we have to check the mode expansion of $T(z)$. Taking the derivative of (\ref{eq:mode-2})
with respect to $z$, we see that the logarithm and the $\xi$ modes vanish. Thus, taking the
normal ordered product $\nop{ T(z) \lid(z)}$ and expanding it by modes yields the same
structure as in (\ref{eq:modelid}). Eventually, some of the modes which vanished for $\lid$
may not vanish for $t(z)$, i.\,e.\ in general the $\hat{l}_n$ may differ from zero, where $\hat{l}_n=l_{n,2}$ in the notation of (\ref{eq:modelog}).

\subsection{Calculations for the bosonic free field construction}
The computation simplifies greatly if we take advantage of known OPEs such as 
$\bra TT \ket$ or $\bra \lid \lid \ket$, taking the ansatz
\begin{eqnarray*}
t(z) t(w) &=& \, \nop{ \eins (z) T(z)}\,\nop{\eins (w) T(w)} \\
&\sim& \, (\eins (z) \eins (w)) (T(z)T(w)) + (\eins(z)T(w))(T(z)\eins(w)) \\
&\quad& + (\eins (z) \eins (w)) \nop{T(z) T(w)} + (T(z)T(w)) \nop{ \eins (z) \eins (w)} \\
&\quad& + (\eins (z) T(w))\nop{ T(z) \eins (w)} + (T(z) \eins (w)) \nop{ \eins(z) T(w)}\, .
\end{eqnarray*}
In this case we have to pay attention carefully to the normal ordering of terms. 
In general, we get a very lengthy expression:
\newpage
\small
\begin{eqnarray*}
&&t(z) t(w)\sim \\
&&\left(4\lambda^2\log^2 (z-w) + \frac{\lambda^2}{2 \alpha_0^2} \lzw - 4\lambda \lzw \eins (w)\right) \frac{\frac{c}{2}}{(z-w)^4}\\
&&+ \frac{\lambda^2}{(z-w)^4} +\frac{\frac{c}{2}}{(z-w)^4}\nop{\eins (w)\eins (w)}\\
&&  -2\lambda\lzw \del \eins (w)\frac{\frac{c}{2}}{(z-w)^3}+\frac{\frac{c}{2}}{(z-w)^3}\nop{\del \eins (w)\eins (w)}\\
&& - \lambda\lzw \del^2 \eins (w)\frac{\frac{1-24\alpha_0^2}{2}}{(z-w)^2}+\frac{\frac{c}{2}}{2(z-w)^2}\nop{\del^2 \eins (w)\eins (w)}+\frac{2\nop{T(w)\eins(w)\eins(w)}}{(z-w)^2} \\
&&+\frac{(\frac{\lambda^2}{\alpha_0^2} \lzw + 8\lambda^2 \log^2 (z-w))T(w)}{(z-w)^2}
- \frac{8\lambda \lzw t(w)}{(z-w)^2}+\frac{2\lambda t(w)}{(z-w)^2} \\
&\quad& + \frac{1}{(z-w)^2} \left( -\del \eins (w) \del \eins(w) - 2\lambda^2 \dpw \dpw + \left(\frac{\lambda^2}{\cons} + \lambda\mathrm{i}a\sqrt{2} \eins(w)\right) \ddpw\right)\\
&&  -\lambda \lzw \del^3 \eins (w)\frac{\frac{c}{2}}{3(z-w)}+\frac{\frac{c}{2}}{6(z-w)}\nop{\del^3 \eins (w)\eins (w)} + \frac{\del \nop{T(w)\eins(w)\eins (w)} }{(z-w)} \\
&& + \frac{(\frac{\lambda^2}{2\alpha_0^2}\lzw+ 4\lambda^2\log^2 (z-w))\del T(w)}{(z-w)}
- \frac{4\lambda \lzw \del t(w)}{(z-w)}+ \frac{\lambda \del t(w)}{(z-w)}\\
&& + \frac{1}{(z-w)} \left(\del \eins(w)\left(-2\lambda \dpw \dpw - \left(\frac{\lambda}{\cons} +\mathrm{i}a\sqrt{2} \eins(w)\right) \ddpw \right) \right.\\
&& \left.+ \left(\frac{\lambda^2}{2\cons} + \frac{\lambda\mathrm{i}a\sqrt{2}}{2}\eins(w)\right) \dddpw - \lambda^2\dpw\ddpw \right)\\
&& -2\lambda\lzw \del^4 \eins (w)\frac{\frac{c}{2}}{24}\\
&& - 2\lambda\lzw \del \eins (w)\del T(w)- \frac{4\lambda}{3} \lzw T(w)\del^2 \eins (w)\\
&& +\left( 4\lambda^2\log^2 (z-w) +\frac{\lambda^2}{2 \alpha_0^2} \lzw - 4\lambda \lzw \eins (w) \right)\\
&& \cdot\left(\frac{1}{2}\nop{\dddpw \dpw} + \frac{\cons}{6} \partial^4 \pw + \nop{T(z)T(w)}\right)\\
&&  + \left(\nop{ \eins (w) \eins (w)} + \frac{\lambda^2}{2 \alpha_0^2} \lzw \right) \frac{\frac{c}{2}}{(z-w)^{4-2a^2}}\\
&&  + \left( \nop{ \del \eins (w) \eins (w)} \right) \frac{\frac{c}{2}}{(z-w)^{3-2a^2}}\\
&&  +  \left( \nop{\del^2 \eins (w) \eins (w)} \right) \frac{\frac{c}{2}}{2(z-w)^{2-2a^2}}
+\frac{\nop{T(w)\eins (w) \eins(w)}}{(z-w)^{2-2a^2}}\\
&&+\frac{\frac{\lambda^2}{\alpha_0^2}\lzw \nop{T(w)\exp(2\mathrm{i}a\sqrt{2}\pw) }}{(z-w)^{2-2a^2}}\\
&&  + \left( \nop{\del^3 \eins (w) \eins (w)} \right) \frac{\frac{c}{2}}{6(z-w)^{1-2a^2}}
+ \frac{\del\nop{ T(w)\eins(w) \eins (w)}\exp(2\mathrm{i}a\sqrt{2})}{(z-w)^{1-2a^2}}\\
&& +\frac{\frac{\lambda^2}{2\alpha_0^2}\lzw \nop{\del T(w)\exp(2\mathrm{i}a\sqrt{2}\pw) }}{(z-w)^{1-2a^2}}\\
&& +\left( \frac{\lambda^2}{2\alpha_0^2}\lzw (z-w)^{-2a^2}\exp(2\mathrm{i}a\sqrt{2}\pw)  +(z-w)^{-2a^2}\nop{ \eins (z) \eins (w)}\right)\\
&& \cdot \left(\frac{1}{2}\nop{\dddpw \dpw} + \frac{\cons}{6} \partial^4 \pw + \nop{T(z)T(w)}\right)\, .
\end{eqnarray*}
\normalsize
Now we can make the assumption $2 a^2 > 4$ ( $a > \sqrt{2}$), 
in order to get rid of the terms proportional to $(z-w)$ to 
some powers of $a$ meaning that no additional fields appear 
in the singular part of the OPE. In this case 
$\alpha_0 > \frac{1}{\sqrt{2}}$, which means that $c=0$ would 
be excluded. But for our $c=-24$ proposal for percolation, it 
is justified. For $c=0$ ($a = \frac{1}{\sqrt{6}}$) we would get 
fractional exponents. Hence the OPE would no longer be valid
for a local chiral field.

\paragraph{A rank three Jordan cell realization}
Based on the augmented $c_{(9,6)} =0$ LCFT, we know that we should
have three $h=0$ fields in the theory, probably belonging to a
rank three Jordan cell structure generated by $L_0$. Thus we take
the simplest possible ansatz for a third field with vanishing
central charge:
\begin{equation}
\hat{\id}(z) = \nop{ \phi \phi }(z).
\end{equation}
Now we can check for its properties, computing the OPE with all
 $h=0$ fields and the stress energy tensor. Obviously,
the OPE with the identity $\id$ has to be trivial.
                                      
Furthermore we have to keep in mind that the identification of 
the two vertex operators corresponding to the solutions of
$h(a) = a(a-2\alpha_0) = 0$ has to hold in both directions.
Hence we get the following OPEs:
\begin{eqnarray*}
T(z) \hat{\id}(w) &=& \frac{\id - 4 \alpha_0^2 \lid(w)}{(z-w)^2} - \frac{\del \hat{\id}(w)}{(z-w)} \, ,\\
\lid(z)\hat{\id}(w) &=& -2 \hat{\id}(w) \lzw - 2 \log^2(z-w) \\
&\quad&- a^2 \lid (w)\lzw - \lid(w)\lzw \, ,\\
\hat{\id}(z)\hat{\id}(w) &=& 2 \log^2(z-w) + 4 \lzw \hat{\id}(w)\, .
\end{eqnarray*}
Obviously, $\hat{\id}$ is a standard logarithmic field.
\newpage
\subsection[$c=-2$ and the fourfold Ising model]{\boldmath{$c=-2$} and the fourfold Ising model}
The OPE of the tensorized $c=0$ model, 
consists of an ordinary CFT part from the sector with $c=+2$ and and LCFT part from the 
$c=-2$ sector. 
The basic features of the special representation of the $c=-2$ theory
are stated in (\ref{eq:mode-2}) and (\ref{eq:anticom-2}).
The contraction rules follow from
\begin{equation}
\ctpztmw = - \lzw.
\end{equation}
To obtain the two-point functions, we make the same ansatz as before.
With the help of the anticommutation relations we find that
$$
- \frac{1}{2}\nop{ \del^2 \eins (w) \eins(w)} - \nop{\del \eins(w) \del \eins(w)} = - \frac{1}{2}\del^2\nop{  \eins (w) \eins(w)} = 0
$$
and
$$
- \frac{1}{3}\nop{ \del^3 \eins (w) \eins(w)} - \del \nop{\del \eins(w) \del \eins(w)} = - \frac{1}{3}\del^3 \nop{ \eins (w) \eins(w)} = 0
$$
as well as
$$
\nop{ \eins (w) \eins(w)} \, \mathrm{\ \ and\ \ }\, \del\nop{ \eins (w) \eins(w)} = 0\, .
$$
Thus the OPE reduces to
\small
\begin{eqnarray*}
&&t(z)t(w) \sim\\
&& \frac{1}{(z-w)^4}\left( \log^2 (z-w) - 2 \lzw \eins (w) + 1 \right)\\ 
&& -\frac{1}{(z-w)^3} \lzw \del \eins(w) \\
&& +\frac{1}{(z-w)^2}\left( \left[\lzw - 2 \log^2 (z-w) \right] T(w) + \left[2 - 4 \lzw \right] t(w)\right)\\
&& -\frac{1}{(z-w)^2}\left( \frac{\lzw}{2}\del^2\eins(w) \right)\\
&& +\frac{1}{2(z-w)}\left( \left[\lzw - 2 \log^2 (z-w) \right] \del T(w) + \left[2 - 4 \lzw \right] \del t(w)\right) \\
&& +\frac{1}{2(z-w)}\left( \del^3 \eins(w)  -\frac{\lzw}{3}\del^3\eins(w)\right)\\
&& +\left( \log^2 (z-w) - 2 \lzw \eins (w) \right) \nop{ T(z)T(w)} \\
&& + \frac{\lzw}{24}\left( \nop{\ddddtpw \tmw} + \nop{\tpw \ddddtmw} \right)\\
&& - \log^2(z-w)\left( \nop{\dtpw \dddtmw} + \nop{\dddtpw \dtmw} \right)\\
&& +2  \lzw \left( \nop{\dtpw \dddtmw} + \nop{\dddtpw \dtmw} \right) \eins(w)\, .
\end{eqnarray*}
\normalsize
\subsection{Calculations for the general tensorized model}
\enlargethispage{1cm}
We take the ansatz
\small
\begin{eqnarray*}
t^{(0)} (z) t^{(0)} (w)
&=& t^{(1)}(z) t^{(1)}(w) + \left(\alpha + \beta \lid(z) \right)\left(\alpha + \beta \lid(w) \right) T^{(2)}(z) T^{(2)}(w)\\
&\sim& \left(T^{(1)}(z)T^{(1)}(w) + \nop{T^{(1)}(z)T^{(1)}(w)}\right)\left(\lid^{(1)}(z)\lid^{(1)}(w)+ \nop{\lid^{(1)}(z)\lid^{(1)}(w)}\right)\\
&\quad& + \left(T^{(1)}(z)\lid(w)+ \nop{T^{(1)}(z)\lid(w)}\right)\left(\lid^{(1)}(z)T^{(1)}(w)+ \nop{\lid^{(1)}(z)T^{(1)}(w)}\right) \\
&\quad& + \left(\alpha^2 + 2 \alpha \beta \lid + \beta^2 \lid(z) \lid (w) \right)T^{(2)}(z) T^{(2)}(w)\, .\\
\end{eqnarray*}
\normalsize
Note that there are no contractions between the two parts; 
the tensorized fields factorize into their respective OPEs.
\begin{eqnarray*}
&&t^0 (z) t^0 (w)\\
&\sim& \left(\frac{\frac{c_{1}}{2}}{(z-w)^4} + \frac{2 T^{(1)} (w)}{(z-w)^2} + \frac{\partial_w T^{(1)} (w)}{(z-w)} + \nop{T^{(1)}(z)T^{(1)}(w)}\right)\\
&\quad& \quad \quad \quad \quad \times \left(\log^2 (z-w) - 2 \lzw \lid (w)+ \nop{\lid^{(1)}(z)\lid^{(1)}(w)}\right)\\
&\quad& +\left(\frac{\id}{(z-w)^2} + \frac{\partial_w \lid (w)}{(z-w)}+ \nop{T^{(1)}(z)\lid(w)}\right)\\
&\quad& \quad \quad \quad \quad \times\left(\frac{\id}{(z-w)^2} - \frac{\partial_w \lid (w)}{(z-w)}+ \nop{\lid^{(1)}(z)T^{(1)}(w)}\right) \\
&\quad& +\left(\alpha^2 + 2 \alpha \beta \lid (w)+ \beta^2 \lid(z) \lid (w) \right)\\
&\quad& \quad \quad \quad \quad \times\left(\frac{\frac{c_{2}}{2}}{(z-w)^4} + \frac{2 T^{(2)} (w)}{(z-w)^2} + \frac{\partial_w T^{(2)} (w)}{(z-w)}+ \nop{T^{(2)}(z)T^{(2)}(w)} \right)\, .
\end{eqnarray*}

After inserting the OPEs (\ref{eq:OPEansatz}) and sorting the terms by order of $(z-w)$ we get
\small
\begin{eqnarray*}
&\sim&  \frac{1}{(z-w)^4} \left[\left(1 + \alpha^2 \frac{c_{2}}{2}\right) - \left(\frac{c_{1}}{2} + \beta^2 \frac{c_{2}}{2}\right)\log^2 (z-w) \right.\\ 
&\quad& \left.+ 2\left(\frac{c_{1}}{2} + \beta^2 \frac{c_{2}}{2}\right) \lzw \lid (w) + \alpha \beta \lid^{(1)} c_{2} + \left(\frac{c_{1}}{2} + \beta^2 \frac{c_{2}}{2}\right)\nop{\lid^{(1)}(z)\lid^{(1)}(w)}  \right]\\
&\quad& +\frac{1}{(z-w)^2} \left[ 
+ 2 t_0 + \alpha \beta \lid^{(1)} T^{(2)}(w) +  \left( T_0(w) - (1-\beta^2) T^{(2)}(w) \right)\nop{\lid^{(1)}(w) \lid^{(1)}(w)} 
\right.\\
&\quad& \left.
2\left( T_0(w) - (1-\beta^2) T^{(2)}(w) \right)\log^2 (z-w) - 4\left( t_0 - \alpha \beta \lid T^{(2)}(w) \right)\lzw 
\right]\\
&\quad& \frac{1}{(z-w)} \left[
\del t_0 +  \frac{1}{2}\alpha \beta \lid^{(1)} \del T^{(2)}(w)  + \del \left(\left(T_0(w) - (1-\beta^2) T^{(2)}(w)\right)\nop{\lid^{(1)}(w) \lid^{(1)}(w)}\right) 
\right.\\
&\quad& \left.+ 
\left(\del T_0(w) - (1-\beta^2) \del T^{(2)}\right)\log^2 (z-w) - 2\left( \del t_0 - \alpha \beta \lid \del T^{(2)}(w) \right)\lzw \lid^{(1)} (w)
\right]\\
&\quad& + \left(\log^2 (z-w) + 2 \lzw \lid^{(1)} (w)\right)\left(\nop{T_0(w)T_0(w)} - (1-\beta^2)\nop{T^{(2)}(w)T^{(2)}(w)}\right)\, .
\end{eqnarray*}
\normalsize
\bibliography{letterbib1}
\end{document}